\documentclass[11pt]{article}
\usepackage{amsmath,amssymb,epsf}
\usepackage{graphicx,subfigure}

\def\ben{\begin{equation}}
\def\een{\end{equation}}
\input amssym.def
\input amssym.tex
\def \mathbb{\Bbb}

\def\crampest{\medmuskip = 1mu plus 1mu minus 1mu}
\def\uncramp{\medmuskip = 4mu plus 2mu minus 4mu}

\def\nn{\nonumber}

\let\bm=\bibitem

\newcommand{\be}{\begin{equation}}
\newcommand{\ee}{\end{equation}}
\def\ba{\begin{array}}
\def\ea{\end{array}}
\def\ft#1#2{{\textstyle{\frac{\scriptstyle #1}{\scriptstyle #2}}}}
\def\fft#1#2{\frac{#1}{#2}}
\def\del{\partial}

\def\sst#1{{\scriptscriptstyle #1}}

\def\td{\tilde}
\def\wtd{\widetilde}
\def\ie{\rm i.e.\ }
\def\dalemb#1#2{{\vbox{\hrule height .#2pt
        \hbox{\vrule width.#2pt height#1pt \kern#1pt
                \vrule width.#2pt}
        \hrule height.#2pt}}}
\def\square{\mathord{\dalemb{6.8}{7}\hbox{\hskip1pt}}}

\newcommand{\hoch}[1]{$\, ^{#1}$}
\newcommand{\bea}{\begin{eqnarray}}
\newcommand{\eea}{\end{eqnarray}}

\def\0{{\sst{(0)}}}
\def\1{{\sst{(1)}}}
\def\2{{\sst{(2)}}}
\def\3{{\sst{(3)}}}
\def\4{{\sst{(4)}}}
\def\5{{\sst{(5)}}}
\def\6{{\sst{(6)}}}
\def\7{{\sst{(7)}}}
\def\8{{\sst{(8)}}}

\def\ep{{\epsilon}}

\def\R{\rlap{\rm I}\mkern3mu{\rm R}}

\def\R{\rlap{\rm I}\mkern3mu{\rm R}}

\def\R{{{\mathbb R}}}

\def\CP{{{\mathbb C}{\mathbb P}}}

\def\HP{{{\mathbb H}{\mathbb P}}}
\def\A8{{{\mathbb A}_8}}

\newcount\hour \newcount\minute
\hour=\time  \divide \hour by 60
\minute=\time
\loop \ifnum \minute > 59 \advance \minute by -60 \repeat
\def\nowtwelve{\ifnum \hour<13 \number\hour:%           % supresses leading 0's
                      \ifnum \minute<10 0\fi%           % so add it it
                      \number\minute
                      \ifnum \hour<12 \ A.M.\else \ P.M.\fi
         \else \advance \hour by -12 \number\hour:%     % supresses leading 0's
                      \ifnum \minute<10 0\fi%           % add it in
                      \number\minute \ P.M.\fi}
\def\nowtwentyfour{\ifnum \hour<10 0\fi%                % need a leading 0
                \number\hour:%                          % supresses leading 0's
                \ifnum \minute<10 0\fi%                 % add it in
                \number\minute}

\begin{document}
\begin{flushright}
DAMTP-2007-4\ \ \ \  MIFP-07-02\\
{\bf hep-th/0701190}\\
January\  2007
\end{flushright}

\vspace{10pt}
\begin{center}

{\Large {\bf Gravitational Solitons and the Squashed Seven-Sphere}}

\vspace{20pt}

P. Bizo\'n$^{1}$, T. Chmaj$^{2,3}$, G.W. Gibbons$^{4}$ and  C.N.
Pope$^{5}$

\vspace{20pt}

{\hoch{1}\it
M. Smoluchowski Institute of Physics, Jagiellonian University,
Krak\'ow, Poland}

{\hoch{2}\it H. Niewodniczanski Institute of Nuclear Physics,~Polish
Academy of Sciences, Krak\'ow,~Poland }

{\hoch{3}\it
Cracow University of Technology, Krak\'ow, Poland
}

{\hoch{4}\it
D.A.M.T.P., Cambridge University, Wilberforce Road,
Cambridge CB3 0WA, UK
}

{\hoch{5}\it George P. \&  Cynthia W. Mitchell Institute
for Fundamental Physics\\
Texas A\&M University, College Station, TX 77843-4242, USA}

\vspace{10pt}

\vspace{40pt}

%$^1$ M. Smoluchowski Institute of Physics, Jagiellonian University,
%Krak\'ow, Poland

%$^2$ H. Niewodniczanski Institute of Nuclear Physics,
%Polish Academy of Sciences, Krak\'ow, Poland

%$^3$Cracow University of Technology, Krak\'ow, Poland

%$^4$ George P. \& Cynthia W. Mitchell Institute for Fundamental Physics,
%Texas A\&M University, College Station, TX 77843-4242, USA

%$^5$ D.A.M.T.P., Cambridge University, Wilberforce Road,
%Cambridge CB3 0WA, UK

\underline{ABSTRACT}
\end{center}

   We discuss some aspects of higher-dimensional gravitational solitons
and kinks, including in particular their stability.  We illustrate our
discussion with the examples of (non-BPS) higher-dimensional
Taub-NUT solutions as the spatial metrics in $(6+1)$ and $(8+1)$
dimensions.  We find them to be stable against small but non-infinitesimal
disturbances, but unstable against large ones, which can lead to black-hole
formation.  In $(8+1)$ dimensions we find a continuous non-BPS family of
asymptotically-conical solitons
connecting a previously-known kink metric with the supersymmetric $\A8$
solution which has Spin(7) holonomy.  All the solitonic spacetimes we
consider are topologcally, but not geometrically, trivial.  In an
appendix we use the techniques developed in the paper to establish the
linear stability of five-dimensional Myers-Perry black holes with
equal angular momenta against cohomogeneity-2 perturbations.

%\centerline{\today\,\,\now}

\tableofcontents

\section{Introduction}

\subsection{Gravitational solitons and kinks}

By analogy with other areas of physics,  a {\it Gravitational Soliton}
in $n$ spacetime dimensions
may be defined to be an everywhere complete non-singular
globally stationary Lorentzian spacetime $M$, satisfying
the vacuum Einstein equations \cite{gibdef}.\footnote{In this paper
we shall be
concerned only with the case of vanishing cosmological constant
although many of the ideas go through in the case that the cosmological
constant is negative}
Thus a gravitational soliton has by definition
neither an event horizon nor an ergo-sphere and should therefore
be distinguished from a stationary or static
black hole, which is only required to be
non-singular outside a a regular event horizon.  For conventional solitons
in flat space, one usually adds as a requirement that the solution
not only be non-singular, but also have finite total energy.  Furthermore,
this energy is determined by the
conserved charges the soliton may carry and also by quantities
specifying asymptotic boundary conditions.  Thus, for example, a
magnetic monopole in Yang-Mills theory has a mass determined by its
magnetic charge and by the vacuum expectation value of the Higgs field
at infinity.

    There are no gravitational solitons without horizons
in four spacetime dimensions
but in higher dimensons such objects do exist: perhaps the best known
example being the Kaluza-Klein monopole in five dimensions
\cite{groper,sork}.
This has a metric of the form
%%%%%%
\be
d\hat s_5^2 =-dt^2 + ds_4^2\,,
\ee
%%%%%
where $ds_4^2$ is the self-dual Taub-NUT gravitational instanton metric.
This four-dimensional metric is asymptotically locally flat, with a
circle direction that stabilises to a constant length at large distance.
The Kaluza-Klein monopole, upon reduction along this circle to $(3+1)$
spacetime dimensions, has a finite ADM mass given by the length of the
circle at infinity.

\subsection{Vacuum interpolation}

      In common with many flat-space solitons, the Kaluza-Klein
monopole may be thought of
as spatially interpolating between two inequivalent ``vacua'' or
``ground states'' of the theory, namely the flat
five-dimensional Minkowski spacetime ${\Bbb E} ^{4,1}$
near the origin, and the compactified Kaluza-Klein ground state
$S^1 \times{\Bbb E} ^{3,1}$ near infinity.
If a solitonic solution of the Einstein equations exhibits vacuum
interpolation, it seems reasonable to refer to it also as a
{\it Gravitational Kink}.\footnote{Note that the use of the word kink
here should be distinguished  from the notion of the ``kink number''
of a Lorentzian metric with respect to some hypersurface, introduced by
Finkelstein and Misner \cite{finmis}, and elaborated upon in \cite{gibhaw}.}

   Again by analogy with other areas of physics, one does not expect
to find an {\it Asymptotically Flat}   gravitational soliton,
i.e. one which outside a compact spatial set or world tube
tends to the flat metric on $n$-dimensional Minkowski spacetime
${\Bbb E} ^{n-1,1}$.   This is because it would interpolate between
two copies of the same vacuum.   Indeed, as we shall see shortly,
this intuitive expectation is borne out by a No-Go theorem.
Thus with respect to the flat Minkowski vacuum
${\Bbb E} ^{n-1,1}$, one cannot think of a gravitational soliton
as having finite energy with respect to the flat Minkowski vacuum.
Nevertheless, with respect to
the ground state near infinity it may well have finite energy.

    Vacuum interpolation also occurs in the case of extreme black holes
or extreme black $p$-branes.  Again, this is between different kinds of ground
states; typically between a flat vacuum at infinity, and a compactified
AdS$_{p+2}\times K$, where $K$ is an $(n-p-2)$-dimensional compact
manifold \cite{gibtow}.

\subsection{Classical stability}

   The definition given above does not
specify whether a gravitational soliton should be stable. That is
deliberate, because  although to be important as a potentially
long-lived classical ``lump,'' to use Coleman's phrase \cite{cole},
a gravitational soliton should at least be be classically stable
against {\sl small} but non-infinitesimal disturbances (i.e. not
merely linearized fluctuations). It is not reasonable, however, to
demand classical  stability, in any gravitational theory, against
all possible {\sl large} disturbances, since nothing forbids
gravitational collapse to a black hole with the same asymptotics.
After all, even Minkowski spacetime is unstable against the
possibility of a concentrated region of gravitational waves
undergoing gravitational collapse to a black hole.  Indeed in  a
recent numerical study of the dynamics of Kaluza-Klein monopoles
\cite{BCG}, precisely such a collapse was seen to occur.  The
possibility of black hole formation, and the resulting spacetime
singularities, invalidate the type of of topological stability
criteria derived from cobordism theory that were discussed in \cite{bagoru}.

   Both Minkowski spacetime and the Kaluza-Klein monopole are
supersymmetric, or BPS.  Thus these two examples clearly demonstrate
the fallacy of the widespread belief that to prove stability in
general relativity it suffices to establish that the spacetime
admits a Killing spinor.\footnote{Quite apart from the non-linear
instabilities
involving black-hole formation, further instabilities of Minkowski
spacetime, or indeed any asymptotically
flat spacetime, can arise unless one restricts attention to perturbations
or deformations that decay at large distances.  The example of Kasner
spacetime clearly shows that flat space is unstable to the formation
of all-encompassing naked singularities in finite time,
if one allows perturbations that do not decay near infinity.}

   The singularity theorems of classical general relativity show that
these black holes contain spacetime singularities, which are a clear
indication that the classical theory is incomplete.  If these
singularities are hidden inside event horizons, i.e. if cosmic
censorship holds, they may not be an obstacle to studying the
evolution of the exteriors of black holes.  However, such black holes
do not have a fixed mass, and may grow by, for example, absorbing
radiation.  Thus in general, black holes cannot be thought of as
solitons.  An exception may arise if one considers so-called extreme
black holes, in which the mass may be determined entirely in terms of
conserved charges \cite{gibbs}.  We shall not discuss this
type of ``solitonic'' black hole further in this paper.

\subsection{Quantum stability}

   The stability considerations described above were purely classical.
Quantum mechanically, black holes are unstable against thermal
Hawking radiation.  Thus in the case of the collapsed Kaluza-Klein
monopole, it seems very likely that the ultimate quantum-mechanical
state is the monopole itself, since the magnetic charge cannot be
radiated away.

   If, as is currently widely believed, the evaporation of neutral
black holes leads to their complete disappearance, it would seem that
Hawking evaporation is {\sl essential} in order to solve the
problem of classical singularities.

   A frequently used criterion for the stability of a particle in
quantum mechanics is that it has the least mass of any state carrying
the same conserved charges.  This criterion, often associated with
BPS bounds, is often used to argue that various solitons, or indeed
ground states, are stable.  In the case of spacetimes, what is often
in one's mind is quantum tunnelling.  While it is certainly true that
a BPS condition, or the existence of a Killing spinor, may mean that
tunnelling is impossible, it does not rule out the sort of classical
instabilities we discussed previously.

\subsection{Ultra-Staticity}

   The main subject of the present paper is the case of gravitational
solitons in nine spacetime dimensions.  This dimension is large
enough to admit a rather richer structure of solitonic solutions
than can be obtained in lower dimensions. Before describing our new
results however, it may be useful to continue the general
discussion, making it a little more precise. In particular we wish
to establish the general result that that a gravitational soliton,
as we have defined  it, must be {\it ultra-static}, i.e. the
(unwarped)  product $ M= {\Bbb R} \times \Sigma$ of time with  a
complete non-singular Ricci flat spatial Riemannian $(n-1)
$-manifold $\Sigma$.

The assumption of global  stationarity means that the
spacetime is an ${\Bbb R}$-bundle over $\Sigma$ the space of orbits
of time translations.
The metric may thus be cast  in the form
%%%%%
\ben
ds^2 = -V^2  (dt +\omega _idx^i)^2 + g_{ij} dx^i dx^j\,
\een
%%%%%
where $i=1,2,\ldots,n-1$, and the everywhere non-vanishing strictly positive
function
$V$ and the Sagnac ${\Bbb R}$-connection $\omega _idx^i$ are
independent of time. The curvature of the Sagnac connection is
given by
%%%%%
\ben
F_{ij}= \partial _i \omega _j - \partial _j \omega _i \,.
\een
%%%%%

   We begin by noting that the vacuum Einstein equations imply that
%%%%%
\ben
\nabla _i \bigl(V^3 F^{ij}  \bigr)=0\,.
\een
%%%%%
Multiplication by $\omega_j$ and integration by parts gives
%%%%%
\ben
\ft12 \int _\Sigma V^3 F_{ij} F^{ij}  \sqrt{g} d^{n-1} \, x
= \int_\infty V^3 \omega _j F^{ij} d \sigma _i.
\een
%%%%%

If the boundary term at infinity vanishes, then
we conclude that
%%%%%
\ben
F_{ij} =0\,.
\een
%%%%%
The Einstein equations then imply
%%%%%
\ben
\nabla ^2 V=0\,.
\een
%%%%%
Thus if $V$ is bounded at   infinity, a
standard argument based on the maximum principle shows that
%%%%%
\ben
V={\rm constant}
\een
%%%%%
The remaining Einstein equation then implies that
the  spatial metric $g_{ij}$ is Ricci flat
%%%%%
\ben
R_{ij}=0.
\een
%%%%%
If we assume that $\Sigma$ is simply connected, or pass to a finite
covering space if it is not, we may set $\omega_i=0$, $V=1$
 and the metric is ultra-static
%%%%%
\ben
ds^2 =-dt^2 + g_{ij} dx ^i dx ^j\,. \label{ultra}
\een
%%%%%

It follows  from (\ref{ultra}) that the question of the  existence
of gravitational solitons reduces completely to that of the
existence of complete Ricci-flat spatial manifolds $\{\Sigma,
g_{ij}\}$. It is known {\cite{witpos} that there are no non-trivial
{\it Asymptotically Euclidean} \footnote {i.e. which tend to the
flat
  metric
on $(n-1)$-dimensional Euclidean space
${\Bbb E}^{n-1}$ outside a compact set}  metrics, and hence
no asymptotically flat gravitational solitons
but  there are plenty of metrics which are {\it Asymptotically
Locally Euclidean}  \footnote {i.e. which tend to the flat
 metric on  ${\Bbb E}^{n-1}/\Gamma\,, \quad \Gamma \subset O(n-1) $
outside a compact set} as well as metrics with much more complicated
asymptotics.

\subsection{BPS solitons}\label{BPSsec}

Among the various possibilities for the Ricci-flat spatial metric,
of particular  interest are those admitting a covariantly constant
spinor. They have reduced holomony, and  if $n<12$, the spinor field
is a Killing spinor of a supergravity theory and the solitons are
thus supersymmetric. The existence of the Killing spinor allows one
to relate the spectrum of the Lichnerowicz operator acting on
symmetric traceless second-rank tensors to the spectrum of other
differential operators.  In this way, one may establish the
linearized stability of solitons with special holonomy.  For
example, a metric with Spin(7) holonomy admits a covariantly
constant self-dual 4-form.  This 4-form may be used \cite{gibpagpop}
to establish a 1-1 correspondence between the spectrum of the
Lichnerowicz operator and the spectrum of the Hodge de-Rham operator
acting on anti-self-dual 4-forms.  Since the spectrum of the latter
is manifestly non-negative, it follows that the Lichnerowicz
operator has no modes of negative eigenvalue, and hence that Spin(7)
solitons are stable at the linearised level.  However, as we
discussed earlier, they will not be stable against deformations
sufficiently large that collapse to black holes takes place. Similar
remarks about linearised stability apply to the other cases of
special holonomy, which are as follows:

\begin{itemize}
\item Ricci-flat K\"ahler or Calabi-Yau with holonomy
$SU(k) \subset SO(2k)$ and thus $n=2k+1$,
\item Hyper-K\"ahler with  holonomy
$Sp(k) \subset SO(4k)$, and thus $n=4k+1$,
\item Holonomy
$G_2  \subset SO(7)$, and thus $n=8$,
\item Holonomy
$Spin(7)   \subset SO(8)$, and thus $n=9$.
\end{itemize}

Explicit complete non-singular metrics are known in all cases. The
easiest examples to construct are cohomogenity one and are {\it
Asymptotically Conical}(AC); they tend to Ricci flat cones over an
Einstein  manifolds which are:

\begin{itemize}
\item Holonomy
$SU(k) \subset SO(2k)$: Einstein-Sasaki,
\item Holonomy
$Sp(k) \subset SO(4k)$: Einstein-Tri-Sasaski,
\item Holonomy $G_2  \subset SO(7)$ : Weak $SU(3)$,
\item Holonomy
$Spin(7)   \subset SO(8)$: Weak $G_2$.
\end{itemize}

Other explicit cohomogenity one examples have been found which are
{\it  Asymptotically Locally Conical} (ALC). In this case
a circle  subgroup of the isometry group has orbits which tend to constant
length at  infinity. The ur-example is the  Taub-NUT metric, i.e. the
Kaluza-Klein monopole.

\subsection{Cohomogeneity One and Cohmogeneity Two}

In this paper we shall restrict attention to
complete  Ricci flat $(n-1)$-dimensional positive definite metrics
which are of cohomogeneity one, that is whose isometry group
$G$ has principal orbits which are $(n-2)$ dimensional.
Such solutions  give rise to static solitions on $n$-dimensional
Lorentzian spacetime obtained by taking the metric product with
time. The isometry group of the $n$-dimensional static
spacetime is therefore the product ${\Bbb R} \times G$.

We then construct a consistent time-dependent ansatz
for the $n$-dimensional spacetime,   which is
is invariant under just the  action of the orginal group
$G$, and which agrees with the static soliton solution
in the special case that there is no time dependence.
The general time-dependent Lorentzian  spacetime is  thus of cohomogeneity
two. The word ``consistent'' means  that every solution
of the resulting system of $(1+1)$-dimensional equations
gives  a solution of the $n$-dimensional vacuum Einstein equations.

The reason for restricting to cohomogeneity one and  cohomogeneity
two is not only for simplicity but because it allows us to exploit
the considerable body of existing information in the literature
on cohomogeneity-one Ricci-flat metrics.

We shall also mainly concentrate on the case when the spatial
manifold is topologically, but not geometrically ${\Bbb R}^n$.

\section{Higher-Dimensional Time-Dependent Taub-NUT Solitons}
\label{Taub-NUTsec}

   One may consider a variety of higher-dimensional static metrics of
the form $d\hat s^2 = -dt^2 + ds^2$, where $ds^2$ is a Ricci-flat
spatial soliton metric.  Then, following the same strategy as in
\cite{BCG}, these
metrics may be used to provide initial data for the time-dependent vacuum
Einstein equations.  Following the discussion in \cite{BCG}, we
shall consider the
higher-dimensional analogues \cite{baibat,pagpop} of the four-dimensional
self-dual Taub-NUT metric.
It should be noted, however, that these higher-dimensional
analogues are non-supersymmetric, in the sense that unlike the
four-dimensional Taub-NUT case, there is no Killing spinor.  We shall
discuss the examples of the Taub-NUT metrics in six and eight dimensions
below, after first presenting a general time-dependent ansatz.

\subsection{The time-dependent ansatz}\label{mredsec}

   A suitably general time-dependent  ansatz for our purposes is
%%%%%
\be
d\hat s^2 = - A e^{-2\delta}\, dt^2 + A^{-1}\, dr^2 +
   r^2 [e^{3B}\, d\Sigma_m^2 + e^{-6mB}\, \sigma^2]\,,
\label{mans}
\ee
%%%%%
where $d\Sigma_m^2$ is the metric on an Einstein-K\"ahler space of
real dimension $2m$, normalized so that it satisfies $R_{ab}= 2(m+1)
g_{ab}$. The 1-form $\sigma$ is given by $\sigma=d\psi+ {\cal B}$,
where ${\cal B}$ is a potential on the Einstein-K\"ahler base space,
such that $d{\cal B}= 2J$, where $J$ is the K\"ahler form.  The
total dimension of the spacetime is $D=2m+3$.

   Special cases of (\ref{mans}) include, if $e^{2\delta}=A$ and the
metric is assumed to be independent of time, the Taub-NUT solitons.
For these solutions, the Einstein-K\"ahler base spaces are taken to
be $\CP^m$. Other important special cases are the higher-dimensional
Kerr and Kerr-AdS metrics.  In the case of $(2m+3)$ dimensions, with
all rotation parameters set equal, these have cohomogeneity one, and
their stability can be analysed by a small extension of our
procedure in which the Kaluza-Klein vector is retained also in the
reduction.  This is discussed in the Appendix.

   The time-dependent Einstein equations
$\hat R_{MN}= 2(m+1)\lambda\, \hat g_{MN}$ for the metric (\ref{mans})
break up into momentum  and Hamiltonian constraint equations
%%%%%
\bea A' &=& -\fft{2m A}{r} + \fft{2m}{(2m+1)r}\, [2(m+1)e^{-3B} -
e^{-6(m+1)B}] -
  \frac{9}{2} m r(e^{2\delta} A^{-1}\, \dot B^2 + A {B'}^2) \nn\\
&&- \fft{c^2}{2(2m+1)}\,r^{-4m-3}\,
     e^{6m B}- 2(m+1)\lambda\, r\,,\nn\\
\dot A &=& - 9 m r A \dot B B'\,,\label{constraintsm} \eea
%%%%%
a slicing condition
%%%%%
\be \delta' = -\frac{9}{2} m r (e^{2\delta} \, A^{-2}\, \dot B^2 +
{B'}^2)\,,\label{slicem} \ee
%%%%%
and a wave equation
%%%%%
\bea
&&\Big( e^\delta\, A^{-1}\, r^{2m+1}\, \dot B\Big)^{.} -
 \Big(e^{-\delta}\, A\, r^{2m+1}\, B'\Big)' + \fft{4(m+1)}{3(2m+1)}
 e^{-\delta}\, r^{2m-1} \,
  (e^{-3B} - e^{-6(m+1)B}) \nn\\
&&+
      \fft{c^2}{2(2m+1)} r^{-2m-3} \, e^{-\delta}\, e^{6mB}=0\,.
\label{wavem}
\eea
%%%%%

   If we define a mass functional $M(r,t)$, by writing
$A=1 - M(r,t)/r^{2m} - \lambda r^2$, then the Hamiltonian constraint becomes
%%%%%
\bea M'&=& \frac{9}{2} m r^{2m+1}\, (e^{2\delta} A^{-1} {\dot B}^2 +
A {B'}^2) +
\fft{c^2}{2(2m+1)} r^{-2m-3}\, e^{6mB}\nn\\
&&
   + \fft{2m}{2m+1}\, r^{2m-1} [(2m+1) + e^{-6(m+1)B} - 2(m+1) e^{-3B}]\,.
\eea
%%%%%
Note that the right-hand side is manifestly positive.

  In what follows, we shall specialise to the case of the vacuum
Einstein equations, by setting $\lambda=0$.

\subsection{The $(6+1)$-dimensional Taub-NUT soliton}

   The spatial metric of the time-independent $(6+1)$-dimensional
Taub-NUT soliton is given by
%%%%%
\be
ds_6^2 = \fft{(\rho+\ell)^2\, d\rho^2}{2(\rho-\ell)(\rho+3\ell)}
 + \fft{2\ell^2 (\rho-\ell)(\rho+3\ell)}{(\rho+\ell)^2}\, \sigma^2 +
  (\rho^2-\ell^2)\, d\Sigma_2^2\,,\label{d6taub}
\ee
%%%%%
with $\rho\ge\ell$, where the notation for $\sigma$ and
$d\Sigma^2$ here is the same as in (\ref{mans}).  One can see that
the metric near $\rho=\ell$ approaches the origin of hyperspherical
coordinates in $\R^6$, by defining a new radial coordinate
$y= \sqrt{2\ell(\rho-\ell)}$.   At large $\rho$, the metric approaches
$\R^5$ times a circle of asymptotic radius $\sqrt2\, \ell$.  The
manifold on which (\ref{d6taub}) is defined has the topology of $\R^6$.

   Comparing with the ansatz (\ref{mans}), with $m=2$, we see that
the six-dimensional Taub-NUT metric gives initial data with
%%%%%
\be e^{-15 B_0} = \fft{2\ell^2(\rho+3\ell)}{(\rho+\ell)^3}\,,\qquad
A_0= \fft{2(\rho+3\ell)(\rho-\ell)}{(\rho+\ell)^2}\,
      \Big(\fft{dr}{d\rho}\Big)^2\,,\qquad e^{2\delta_0}= A_0\,,
\ee
%%%%%
where
%%%%%
\be
r^{10} = 2\ell^2 (\rho-\ell)^5 (\rho+\ell)^2(\rho+3\ell)\,.
\ee
%%%%%

\subsection{The $(8+1)$-dimensional Taub-NUT soliton}

   The spatial metric of the time-independent $(8+1)$-dimensional
Taub-NUT soliton is given by
%%%%%
\be
ds_8^2 = \fft{5(\rho+\ell)^3\, d\rho^2}{8(\rho-\ell)(\rho^2 + 4\ell \rho
    + 5\ell^2)} +
 \fft{8\ell^2(\rho-\ell)(\rho^2 + 4\ell \rho + 5\ell^2)}{5(\rho+\ell)^3}
\, \sigma^2 + (\rho^2-\ell^2)d\Sigma_3^2\,,\label{d8taub}
\ee
%%%%%
with $\rho\ge\ell$.  It is defined on the manifold $\R^8$.

   Comparing with (\ref{mans}) with $m=3$, we see that the eight-dimensional
Taub-NUT metric gives initial data with
%%%%%
\be e^{-21 B_0}= \fft{8\ell^2(\rho^2+4\ell\rho +
5\ell^2)}{5(\rho+\ell)^4}\,, \qquad A_0=
\fft{8(\rho-\ell)(\rho^2+4\ell\rho+5\ell^2)}{5(\rho+\ell)^3}\,
\Big(\fft{dr}{d\rho}\Big)^2\,,\qquad e^{2\delta_0}=A_0\,, \ee
%%%%%
where
%%%%%
\be
r^{14} = \ft85\, \ell^2 (\rho+\ell)^3(\rho-\ell)^7
      (\rho^2+4\ell\rho + 5\ell^2)\,.
\ee
%%%%%
As we shall see below this solution is a special case of a more
general ansatz.

\section{Nine-Dimensional Squashed Seven-Sphere Solitons}

   Our previous ansatz (\ref{mans}) was based on the Hopf fibring of
$S^{2m+1}$ by $U(1)$ Hopf fibres over a $\CP^m$ base manifold\footnote{Or
fibrations over more general Einstein-K\"ahler base manifolds.}.
In the case where $m=2p$, one can instead consider $S^{4p+3}$ regarded
as an $SU(2)$ bundle over $\HP^p$.  The simplest case is for $p=1$, with
$S^7$ regarded as an $SU(2)$ bundle over $S^4$.  In this section, we
shall consider a time-dependent ansatz for an $(8+1)$-dimensional
time-dependent metric where the spatial 8-metric has surfaces at constant
radius that have $S^7$ topology, fibred by $S^3$.  Two deformation parameters
will be included, one parameterising the volume of the $S^3$ fibres, and
the other parameterising a homogeneous squashing of the $S^3$ fibres
themselves.

   We begin with some group-theoretic preliminaries, by considering
left-invariant 1-forms $L_{AB}$ for $SO(5)$,  These obey
$L_{AB}=-L_{BA}$ and
%%%%%
\be
dL_{AB}= L_{AC}\wedge L_{CB}\,.
\ee
%%%%%
We take the $SO(5)$ indices to range over $0\le A\le 4$, and split them as
$A=(a,4)$, with $0\le a\le 3$.  The $SO(4)$ 1-forms $L_{ab}$ are then
expressed in an $SU(2)_L\times SU(2)_R$ basis with generators
%%%%%
\be
R_i= \ft12 (L_{0i} + \ft12 \ep_{ijk} L_{jk})\,,\qquad
L_i= \ft12 (L_{0i} - \ft12 \ep_{ijk} L_{jk})\,,
\ee
%%%%%
where $1\le i\le 3$.  The $S^7= SO(5)/S0(3)$ coset is then spanned by the
1-forms
%%%%%
\be
R_i\,,\qquad P_a\equiv \ft12 L_{a4}\,.
\ee
%%%%%
(Note a rescaling of $P_a$, relative to \cite{spin7}.  This is done
for convenience, to avoid factors of 2 later.)  The algebra of the 1-forms
is easily seen to be
%%%%%
\bea
dR_i &=& -\ep_{ijk} R_j\wedge R_k - J^i_{ab}\, P_a\wedge P_b\,,\nn\\
dP_a &=& J^i_{ab}\, R_i\wedge P_b + \wtd J^i_{ab}\, L_i\wedge P_b\,,\\
dL_i &=& \ep_{ijk} L_j\wedge L_k - \wtd J^i_{ab}\, P_a\wedge P_b\,,\nn
\eea
%%%%%
where we have defined antisymmetric self-dual and anti-self-dual
'tHooft tensors $J^i_{ab}$ and $\wtd J^i_{ab}$ by
%%%%%
\bea
J^i_{0j} &=& \delta^i_j\,,\qquad J^i_{jk} = \ep_{ijk}\,,\nn\\
\wtd J^i_{0j} &=& \delta^i_j\,,\qquad \wtd J^i_{jk} =- \ep_{ijk}\,.
\eea
%%%%%
   The metric on the unit round $S^7$ is given by
%%%%%
\be
d\Omega_7^2 = R_i^2 + P_a^2\,.
\ee
%%%%%

   Our general time-dependent ansatz is
%%%%%
\be d\hat s_9^2 = -A\, e^{-2\delta}\, dt^2 + A^{-1}\, dr^2  +r^2\,
\Big\{ e^{-4B}\, [e^{2C}\,(R_1^2+R_2^2) + e^{-4C}\, R_3^2]
  + e^{3B}\, P_a^2\Big\}\,.\label{doublesquash}
\ee
%%%%%
    Straightforward calculations show that the Ricci-flatness of $d\hat s_9^2$
implies the following equations.  First, we have the Hamiltonian and momentum
constraints
%%%%%
\bea A' &=& -\fft{6}{r} A - \fft{6r}{7} A \Big[ \fft{7}{2}({B'}^2 +
 e^{2\delta}\, A^{-2}\, {\dot B}^2 ) +
     ({C'}^2 + e^{2\delta}\, A^{-2}\, {\dot C}^2)\Big]\nn\\
&& +
 \fft{2}{7r}\, \Big[ - e^{4B-8C}\! +\! 4e^{4B -2C}\! -\!
    4 e^{-10B + 2C}\! -\! 2 e^{-10B - 4C}\! +\!
      24 e^{-3B}\Big]\,,
      %\nn
      \\
\dot A &=& - 6 r\, A
   ( \dot B B' + \fft{2}{7} \dot C C')\,.
\eea
%%%%%
In addition, there is the slicing constraint
%%%%%
\be \delta' = -3 r\, \Big[
    ({B'}^2 + e^{2\delta}\, A^{-2}\, {\dot B}^2) + \fft{2}{7}
({C'}^2 + e^{2\delta}\, A^{-2}\, {\dot C}^2)\Big]\,. \ee
%%%%%
Finally, we have the dynamical equations for the two squashing modes,
which give
%%%%%
\bea &&e^{\delta}(e^\delta\, A^{-1}\, r^7 \dot B)\dot{} -
   e^{\delta}(e^{-\delta}\, A\, r^7 B')'
\nn\\
&&-
  \fft{4r^5}{21}\, ( - e^{4B - 8C}\! +\! 4 e^{4B - 2C}
   \!+\! 10 e^{-10B + 2C}\! +\! 5 e^{-10B - 4C}\! -\!
  18 e^{-3B})=0\,,
  %\nn
  \\
&&\nn\\
&& e^{\delta}(e^\delta\, A^{-1}\, r^7 \dot C)\dot{} -
   e^{\delta}(e^{-\delta}\, A\,r^7 C')' \nn\\
&&+\fft{4 r^5}{3}\, (-e^{4B - 8C}+
  e^{4B - 2C} + e^{-10B +2C} -
   e^{-10B - 4C})=0\,.
\eea
%%%%%
It can be straightforwardly verified that the constraints are indeed
consistent with the dynamical equations of motion.

    As a check, it can be verified that if we set  $C=
\ft72 B$, for which the ansatz (\ref{doublesquash}) reduces to the
special case of setting $m=3$ and ${\cal A}=0$ in (\ref{mans}), \ie
describing a squashing of $S^7$ viewed as a $U(1)$ bundle over
$\CP^3$, we indeed obtain the same equations as those given in
section \ref{mredsec}. Another check is instead to set $C=0$, in
which case the system reduces to the one discussed in \cite{BCS2},
where $S^7$ is viewed as a round $S^3$ bundle over $S^4$.

\subsection{Static solutions}

In this section we consider regular static asymptotically
(locally) conical solutions of the system (29-33).
Note that in agreement with
Section~1.5 all static solutions are ultrastatic, i.e.,
$Ae^{-2\delta}=1$, thus equations (29-33) reduce to the following
system of ordinary differential equations
%%%%%
\be
   e^{\delta}(e^{\delta}r^7 B')'+
  \fft{4r^5}{21}\, ( - e^{4B - 8C}\! +\! 4 e^{4B - 2C}
   \!+\! 10 e^{-10B + 2C}\! +\! 5 e^{-10B - 4C}\! -\!
  18 e^{-3B})=0, \label{brstat}
\ee
\be
   e^{\delta}(e^{\delta} r^7 C')'
-\fft{4 r^5}{3}\, (-e^{4B - 8C}+
  e^{4B - 2C} + e^{-10B +2C} -
   e^{-10B - 4C})=0\,. \label{crstat}
\ee  \be \delta' = -3 r\, (
    {B'}^2 + \fft{2}{7}
{C'}^2)\,. \label{deltarstat}\ee
%%%%%
 Regularity at the origin implies the
following behavior for $r\rightarrow 0$
 \be B \sim b r^2, \qquad C \sim c r^2,\qquad
e^{\delta}=1-(3b^2+\frac{6}{7}c^2)r^4\,, \label{localr}\ee where $b$
and $c$ are free parameters. Using scaling symmetry, without loss of
generality, we can set $b=1$. Then, (\ref{localr}) gives rise to a
unique one-parameter family of local solutions parametrized by $c$.
Numerical analysis shows that for any $c$ in the interval $0 \leq c
\leq 7/2$, these local solutions can be continued to infinity and
thus give the desired global solutions (see Fig.~1).  For the values
of $c$ outside this interval  the solutions become singular for a
finite $r$. The asymptotic behavior of global solutions near
infinity depends on $c$: for $0 < c \leq 7/2$ the squashing modes
grow logarithmically and $\delta$ goes to minus infinity, while for
$c=0$ both $B$ and $\delta$ have finite limits.
Thus unless $c=0$, the space sections are asymptotically locally
conical,  ALC, but in the limiting $c=0$ case
they become asympototically conical (AC).
For two values of $c$ the solutions are known is closed
form: these are  the Taub-NUT solution (\ref{d8taub}) which  corresponds to
$c=7/2$ and the so called $\A8$ solution \cite{spin70,spin7}
which corresponds to $c=2$.

Below we discuss in detail the structure of static solutions and
their stability properties.
\begin{figure}[h] \centering
\includegraphics[width=0.8\textwidth]{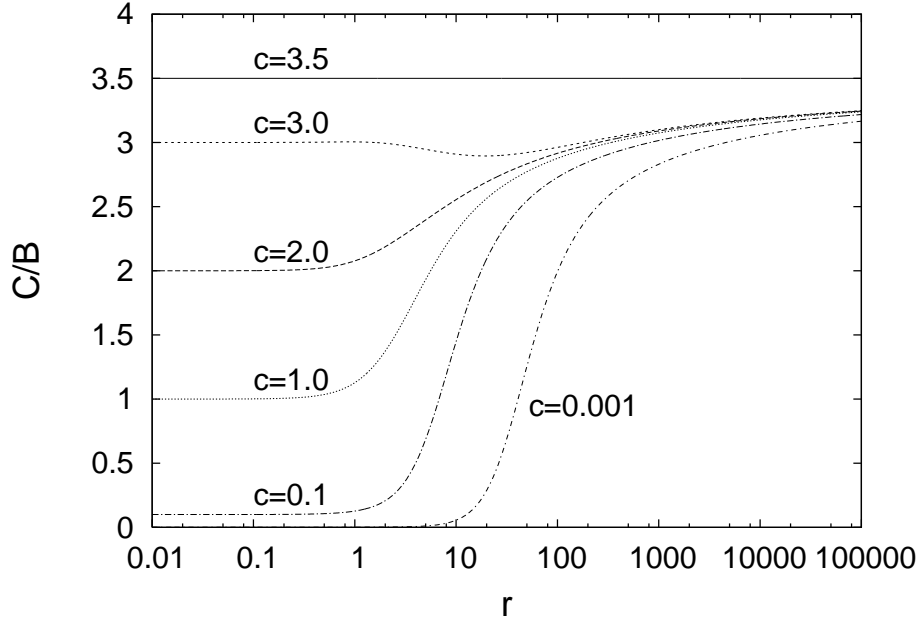}
\caption{\small{Plot of $C/B$ for  static solutions for several
values of the shooting parameter $c$. For $0<c<7/2$ we have $C \sim
(7/2)B-(1/2)\ln{2}$ for $r\rightarrow \infty$. } }\label{fig1}
\end{figure}

\subsection{The Spin(7) background $\A8$ ($c=2$)}

The $\A8$ solution is an ultra-static nine-dimensional  vacuum
solution whose space sections are asymptotically conical (AC).
%%%%%
\be ds_9^2 = -dt^2 + ds_8^2\,,\label{9met} \ee
%%%%%
 where
$ds_8^2$ is a Ricci-flat metric with Spin(7) holonomy.  A simple
metric of this kind, which extends smoothly onto a manifold of
$\R^8$ topology, was obtained in \cite{spin70,spin7}, where it was
denoted by $\A8$.  In the normalisation we are using here, it is
given by
%%%%%
\be ds_8^2 = \fft{(\rho+\ell)^2 d\rho^2}{(\rho+3\ell)(\rho-\ell)} +
   (\rho+3\ell)(\rho-\ell)(R_1^2+R_2^2) +
\fft{4\ell^2(\rho+3\ell)(\rho-\ell)}{(\rho+\ell)^2}\, R_3^2 +
 2(\rho^2-\ell^2) P_a^2\,.\label{A8met}
\ee
%%%%
The radial coordinate $\rho$ lies in the range $\ell\le
\rho\le\infty$. Near $\rho=\ell$ we may define a new radial
coordinate $y=2\sqrt{(\rho-\ell)\ell}$, in terms of which the metric
approaches
%%%%%
\be ds_8^2 \sim dy^2 + y^2 (R_i^2 + P_a^2) \ee
%%%%%
at small $y$.  At large distance, $\rho\rightarrow\infty$, the
metric approaches approaches $\R^7$ times a circle of asymptotic
radius $2\ell$.  The situation is therefore closely analogous to
that of the self-dual Taub-NUT metric in four dimensions.

   Expressed in terms of the ansatz (\ref{doublesquash}), the $\A8$
   solution has the form
%%%%%
\bea e^{21B} &=& \fft{2(\rho+\ell)^5}{\ell^2(\rho+3\ell)^3}\,,\qquad
 e^{-3C} = \fft{2\ell}{(\rho+\ell)}\,,\nn\\
A &=& \fft{(\rho+3\ell)(\rho-\ell)}{(\rho+\ell)^2}\,
    \Big(\fft{dr}{d\rho}\Big)^2\,,\qquad e^{2\delta}= A\,,
\eea
%%%%%
where
%%%%%
\be r^{14} = 64 \ell^2 (\rho+3\ell)^3(\rho-\ell)^7 (\rho+\ell)^2\,.
\ee
%%%%%
\subsection{The continuous family of
solutions ($0<c\leq 7/2$)} We define the new independent variable
$\tau$ by $r e^{\delta} d/dr = d/d\tau$ and let $x=B$ and
$y=\sqrt{\tfrac{2}{7}}\, C$. Then, assuming staticity, equations
(\ref{brstat}) and (\ref{crstat}) take the form
\be \frac{d^2 x}{d\tau^2} + 6 e^{\delta} \frac{d x}{d\tau}+
\frac{\partial V}{\partial x}=0\,, \quad \frac{d^2 y}{d\tau^2} + 6
e^{\delta} \frac{d y}{d\tau} + \frac{\partial V}{\partial y} =0\,,
\label{bctau} \ee
where \be V(x,y)=\fft{1}{21}\Big(- e^{4x-4\sqrt{14}y} + 4
e^{4x-\sqrt{14}y} -4 e^{-10x+\sqrt{14}y} -2 e^{-10x-2\sqrt{14}y}+ 24
e^{-3x}\Big)\,, \label{pot} \ee
and the slicing constraint (\ref{deltarstat}) becomes \be
\frac{d}{d\tau} e^{\delta} = -3 \left(\Big(\frac{d
x}{d\tau}\Big)^2+\Big(\frac{d y}{d\tau}\Big)^2\right)\,.
\label{deltatau}\ee
The boundary conditions (\ref{localr}) translate to the following
asymptotic behavior for $\tau \rightarrow -\infty$
 \be x \sim b\, e^{2\tau},\quad y \sim \sqrt{\frac{2}{7}} c \,e^{2\tau},\quad e^{\delta} \sim
 1-3(b^2+\frac{2}{7} c^2) e^{4\tau}\,. \label{localtau}
 \ee
As above we set $b=1$, hence we have a one-parameter family of local
solutions parametrized by $c$.

It is useful to interpret the above system in terms of the
mechanical analogy of a sticky ball rolling on the surface
$z=V(x,y)$. Due to the friction the energy of the ball,
 \be
    E=\frac{1}{2} \Big[\Big(\frac{d
x}{d\tau}\Big)^2+\Big(\frac{d y}{d\tau}\Big)^2\Big] + V(x,y)\,, \ee
decreases in time \be \frac{d E}{d\tau} = -6 e^{\delta}
\left(\Big(\frac{d x}{d\tau}\Big)^2 + \Big(\frac{d
y}{d\tau}\Big)^2\right) \leq 0\,. \label{lyap}\ee Note that the
combinations of equations (\ref{deltatau}) and (\ref{lyap}) together
with the boundary conditions (\ref{localtau}) yield the constraint
$E=e^{2\delta}$ which can be used to eliminate $e^{\delta}$ from
equations (\ref{bctau}).

Assuming that $x$ and $y$ are both positive and large we can solve
the equations (\ref{bctau})-(\ref{deltatau}) asymptotically to get
the following two possibilities \be \text{(i)} \quad y\sim
\sqrt{\frac{7}{2}} x \qquad \mbox{or} \qquad \text{(ii)} \quad y\sim
\sqrt{\frac{7}{2}} x-\frac{\ln{2}}{\sqrt{14}} \,. \ee The  case (i)
is exceptional and corresponds to the ball rolling down the ridge,
while the case (ii) is generic and it corresponds to the motion down
the valley (on the left side of the valley there is the ridge which
separates it from a cliff and on the right side there is a steep
ascent). In the second case the asymptotic behavior of solutions for
$\tau\rightarrow \infty$ is
 \be x \sim \fft{2}{3}
\ln{\tau} - \fft{1}{3}\ln\left(\fft{245}{10}\right),\qquad
e^{\delta} \sim \frac{6}{\tau}\,. \label{asym_a8} \ee
\begin{figure}[h!]
\centering
\includegraphics[width=0.8\textwidth]{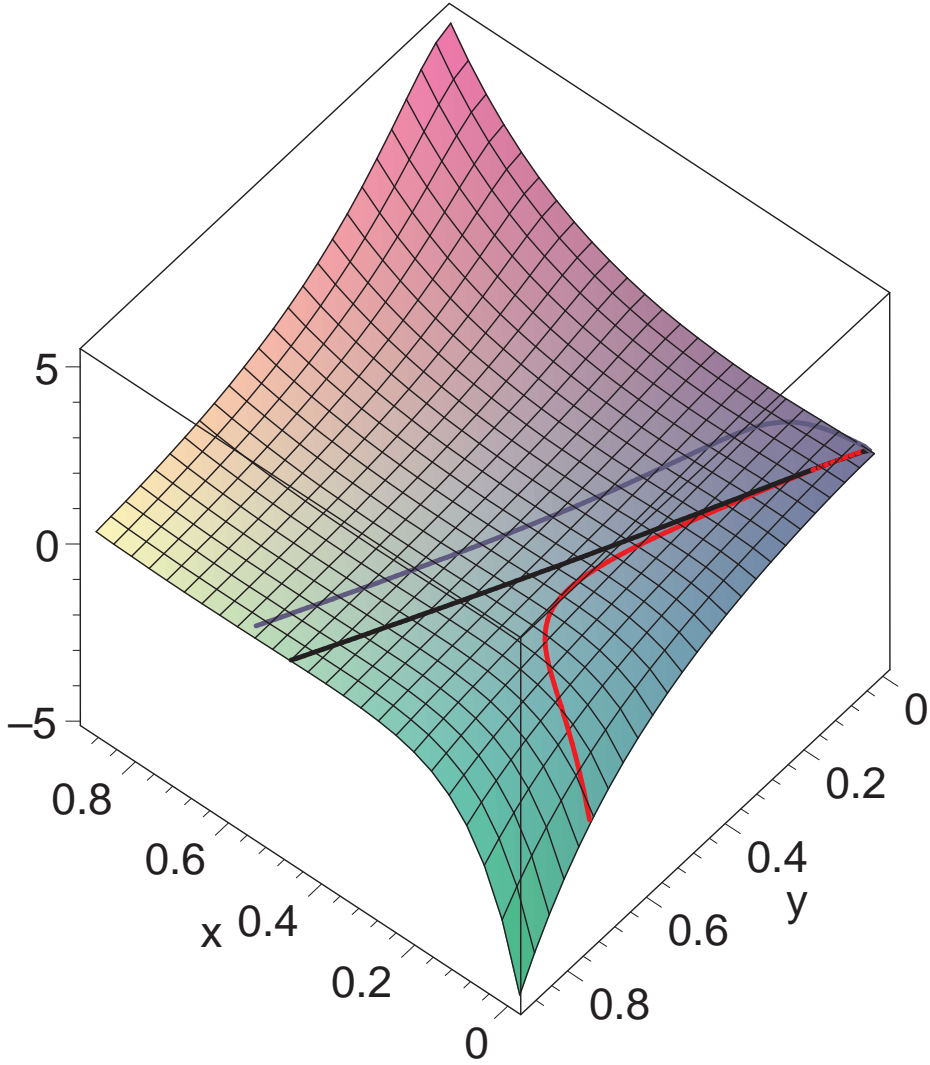}
\caption{\small{The potential $z=V(x,y)$ and three trajectories
which start at the peak $(1,0,0)$ with different slopes (that is,
different values of the parameter $c$).  The black curve ($c=7/2$)
represents the Taub-NUT solution, the dark blue curve ($c<7/2$)
represents a generic solution of the continuous family, and the red
curve shows a singular trajectory with $c>7/2$.}
}\label{fig2}
\end{figure}

\subsection{Stability of static solutions}
The role of static solutions in dynamics depends on their stability
properties. In this section we investigate the stability of static
solutions presented above ($0<c\leq 7/2$).
\subsubsection{Linear stability}
 Following the standard procedure we seek solutions  in the form
\bea
    &B(t,r)= B_0(r)+B_1(t,r), \quad C(t,r)=C_0(r)+C_1(t,r),\\
    &A(t,r)=A_0(r)+A_1(t,r), \quad
    \delta(t,r) = \delta_0(r)+\delta_1(t,r)\,,
\eea where the index $0$ denotes a static solution and the index $1$
denotes a perturbation.
 We substitute this expansion into equations (29)-(33) and linearize them.
Integrating equation (30) we obtain \be A_1=-6 r\,A_0
(B_0'B_1+\fft{2}{7} C_0' C_1)\,, \label{a1} \ee and from equation
(31) we get \be \delta_1'=-6 r (B_0'B_1'+\fft{2}{7} C_0' C_1')\,.
\label{delta1} \ee Inserting equations (\ref{a1}) and (\ref{delta1})
into the linearized equations (32) and (33)
  and separating
the time dependence $B_1(t,r)=\exp(-i \lambda t) v_{\lambda} (r)$,
$C_1(t,r)=\exp(-i \lambda t) u_{\lambda} (r)$ we get the eigenvalue
equation for the spectrum of small perturbations\be -\frac{1}{r^7}
e^{\delta_0}\Big(e^{\delta_0} r^7 \mathbf{S}_{\lambda}\Big)'+
\mathbf{K} \mathbf{S}_{\lambda}= \lambda^2 \mathbf{S}_{\lambda}\,,
\label{stabcont} \ee
 where
\be \mathbf{S}_{\lambda}=\begin{pmatrix} v_{\lambda}\\ u_{\lambda}
\end{pmatrix} \ee and $\mathbf{K}$ is a $2\times2$ matrix
determined by the static solution. The matrix $\mathbf{K}$ is very
complicated but fortunately we do not need it in an explicit form.
To demonstrate stability we exploit the existence of the zero mode
 \be \mathbf{S}_0=r \begin{pmatrix} B_0'(r)\\
C_0'(r)\end{pmatrix}\,, \ee which is due to the scaling invariance
of the problem. From the above heuristic analysis of the behavior of
static solutions it is clear (although not proved rigorously) that
the functions $B_0(r)$ and $C_0(r)$ are monotone increasing, hence
the components of the zero mode have no zeros. Thus, it follows from
the Sturm-Liouville theory that there are no negative eigenvalues.

    We point out that in the case of the $\A8$ manifold, the linearized stability
    of this solution follows from the
existence of a covariantly-constant spinor in the Spin(7) holonomy
manifold $\A8$, as discussed in section \ref{BPSsec}.  Specifically,
the spectrum of the Lichnerowicz operator describing transverse
traceless metric perturbations is identical to the spectrum of the
Hodge-de Rham operator acting on anti-self-dual 4-forms
\cite{gibpagpop}, and thus there can be no negative-eigenvalue modes
and hence the solution is stable at the level of linearized
perturbations.
\subsubsection{Nonlinear stability}
In order to verify numerically the nonlinear stability of static
solutions numerically, we have expressed equations (29)-(33) in the
first order form using the momentum variables \be
 P_B = e^{\delta} A^{-1} \dot B, \qquad P_C = e^{\delta} A^{-1} \dot C\,.
\ee
We have solved the resulting equation system using the free
evolution scheme in which the function $A(t,r)$ is updated from the
momentum constraint (30). Integration in time is done by the
modified predictor-corrector McCormack method on a uniform spatial
grid. The  slicing constraint (31) is solved with the fourth order
Runge-Kutta method. The whole procedure is second order accurate in
time and fourth order in space.
 The
results shown below were produced for initial data of the form
\begin{equation}\label{idata}
 \!   B(0,r)=B_0(r),\quad C(0,r)=C_0(r), \quad P_B(0,r)=p \left(\frac{r}{R}\right)^4
    e^{-\frac{(r-R)^4}{s^4}},\quad P_C(0,r)=0\,,
\end{equation}
where the amplitude $p$ was varied and the parameters $R$ and $s$
were kept fixed. We have found that for small perturbations, that is
for small values of the control parameter $p$, the solution returns
to equilibrium and the excess energy of the perturbation is radiated
away to infinity, while for large perturbations a black hole forms.
The behavior is qualitatively the same for all static solutions
(independently of $c$) and we illustrate it in Figs.~3 and 4
 in the case of the $\A8$ solution.
\begin{figure}[h!]
\centering
\includegraphics[width=0.8\textwidth]{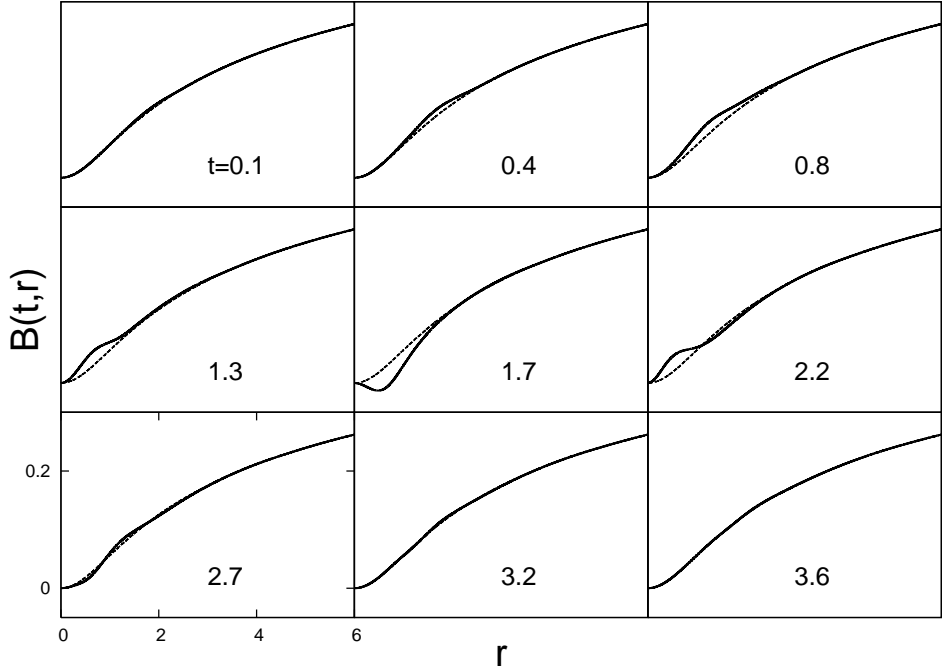}
\caption{\small{Asymptotic stability of the solution $\A8$. For
initial data (\ref{idata}) with a small amplitude ($p=0.1, R=3,
s=1$) we plot a series of snapshots of the function $B(t,r)$ (where
$t$ is central proper time). The dashed line shows the unperturbed
$\A8$ solution.} }\label{fig3}
\end{figure}

\begin{figure}[h!]
\centering
\includegraphics[width=0.8\textwidth]{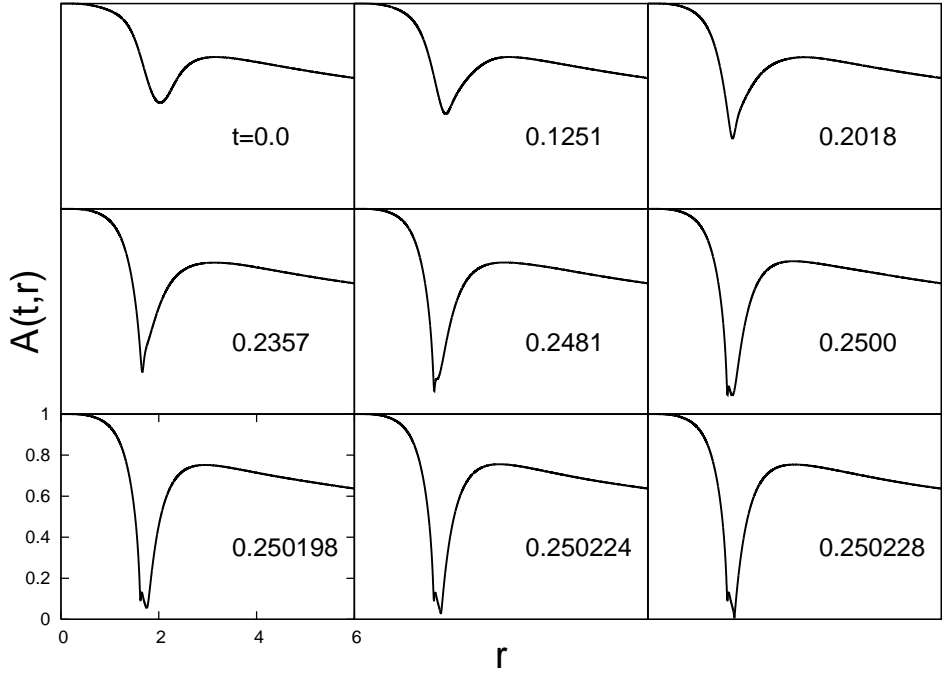}
\caption{\small{Instability of the solution $\A8$ for large
perturbations. For initial data  (\ref{idata}) with a large
amplitude ($p=0.3, R=3, s=1$) we plot a series of snapshots of the
function $A(t,r)$ (where $t$ is  central proper time). During the
evolution $A(t,r)$ drops to zero at $r=r_H \approx 1.773$ which
signals the formation of an apparent horizon there. Outside the
horizon the solution relaxes to a static black hole.} }\label{fig4}
\end{figure}

\subsection{The Eight-Dimensional Kink Solution ($c=0$)}

   In this section, we shall show the existence of a complete
Ricci-flat 8-metric, which is asymptotically conical (AC), which we
shall call the kink.  It may be considered as spatially
interpolating between flat Euclidean 8-space near the origin, \be
B=0,\quad C=0,\quad A=e^{2\delta}=1\,, \label{flatcase} \ee
 and the
Ricci-flat cone over the Einstein-squashed 7-sphere at infinity,
%%%%%
\be B=\frac{\ln{5}}{7}\,,\quad C=0,\quad A=
e^{2\delta}=9\cdot5^{-10/7}\,.\label{conecase} \ee
%%%%%%
   It was demonstrated in \cite{pagpop2}, at the numerical level,
that there exists a complete and non-singular metric that
interpolates between these two constant solutions.  It is defined on
a manifold of $\R^8$ topology.  Here, we present a sketch of a proof
of the existence of this kink solution.

Repeating the steps from Section~3.3 and setting $y=0$ in equations
(\ref{bctau})-(\ref{deltatau}), we obtain the 3-dimensional
autonomous system
%%%%%
\be \frac{d^2 x}{d\tau^2} + 6 e^{\delta} \frac{d x}{d\tau}
+\frac{\partial V}{\partial x}=0\,, \label{bkink}\ee
\be \frac{d}{d\tau} e^{\delta} = -3
\Big(\frac{dx}{d\tau}\Big)^2\,,\label{deltakink} \ee
%%%%%
where \be V=\fft{1}{7}\Big( 8 e^{-3x} + e^{4x} -
    2 e^{-10x}\Big)\,. \label{potkink}
\ee
%%%%%
The boundary conditions (\ref{localr})  for $\tau \rightarrow
-\infty$ simplify to
 \be x \sim b\, e^{2\tau},\qquad e^{\delta} \sim
 1-3b^2 e^{4\tau}\,, \label{localtaukink}
 \ee
hence up to scaling given by $b$ we have a unique regular local
solution.

 As above, we can interpret this system
in terms of the mechanical analogy of a ball rolling in the
potential $V(x)$ with a variable friction (see Fig.~5).
\begin{figure}[h!]
\centering
\includegraphics[width=0.8\textwidth]{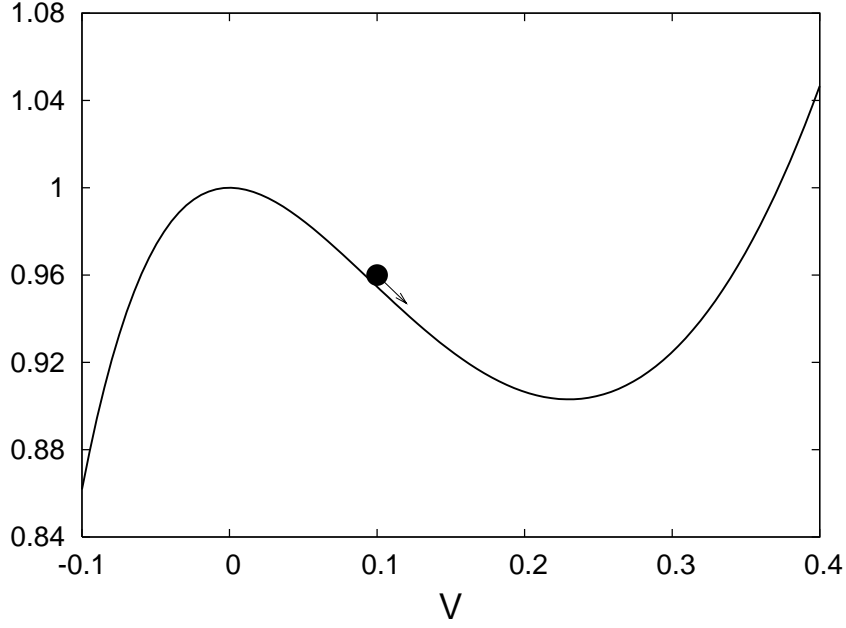}
\caption{\small{In terms of the mechanical analogy the kink solution
corresponds to the ball rolling down from the maximum of the
potential $V=1$ at $x=0$ to the minimum $V=9\cdot 5^{-10/7} \approx
0.903$ at $x=\tfrac{\ln{5}}{7}$.} }\label{fig5}
\end{figure}
The energy of the ball
%%%%%
\be
    E=\frac{1}{2} \Big(\frac{d
x}{d\tau}\Big)^2 + V(x) \ee \label{energykink}
%%%%%
decreases in time because
%%%%%
\be \frac{d E}{d\tau} = -6 e^{\delta} \Big(\frac{d x}{d\tau}\Big)^2
\leq 0\,. \ee
%%%%%
%
As in the general case we have the  constraint $E=e^{2\delta}$.
Using this constraint we eliminate $e^{\delta}$ from equation
(\ref{bkink}) and get the autonomous 2-dimensional dynamical system.
This system has two critical points which correspond to the constant
solutions (\ref{flatcase}) and (\ref{conecase}): the saddle
$(x=0,dx/d\tau=0)$ and the stable node
$(x=\tfrac{\ln{5}}{7},dx/d\tau=0)$.
The function $e^{2\delta}$ serves as the Lyapunov function, thus it
is evident that the orbit starting from the saddle $(0,0)$ along the
unstable manifold will end up at the stable node
$(\tfrac{\ln{5}}{7},0)$ for $\tau \rightarrow \infty$. The
linearization around this critical point yields the following
asymptotic behavior for $\tau \rightarrow
\infty$ \be \begin{pmatrix} x(\tau)-\frac{\ln{5}}{7}\\
dx/d\tau\end{pmatrix} = c_1 e^{\lambda_1 \tau} \mathbf{\xi}_1 + c_2
e^{\lambda_2 \tau}\mathbf{\xi}_2\,, \label{lincrit}\ee where the
eigenvalues are \be \lambda_1=-\frac{8}{5}\cdot 5^{2/7},\qquad
\lambda_2=-2\cdot 5^{2/7}\,,\ee and the corresponding eigenvectors
are
 \be \mathbf{\xi}_1=\begin{pmatrix} 1\\
\lambda_1\end{pmatrix},\qquad  \mathbf{\xi}_2=\begin{pmatrix} 1\\
\lambda_2\end{pmatrix}\,.\ee
The kink orbit approaches the node $(\tfrac{\ln{5}}{7},0)$  along
the slow eigendirection $\mathbf{\xi}_1$. We claim that the kink
orbit stays for all times  in the first quadrant of the
$(x,dx/d\tau)$ plane. To see this, consider an exceptional
trajectory which approaches the node $(\tfrac{\ln{5}}{7},0)$ along
the fast eigendirection $\mathbf{\xi}_2$ (that is, $c_1=0$ and
$c_2<0$ in equation (\ref{lincrit})). This trajectory, run backwards
in time from the node, obviously cannot cross the line
$x=\tfrac{\ln{5}}{7}$ and consequently it prevents the kink
trajectory (which is trapped below the fast eigendirection
trajectory) to do so. In terms of the mechanical analogue of a ball
rolling in the potential $V$ the above analysis demonstrates that
the motion of the ball is overdamped and moreover the ball rolling
down from the maximum of the potential at $x=0$ cannot overshoot the
minimum at $x=\tfrac{\ln{5}}{7}$.

\begin{figure}[h!]
\centering
\includegraphics[width=0.8\textwidth]{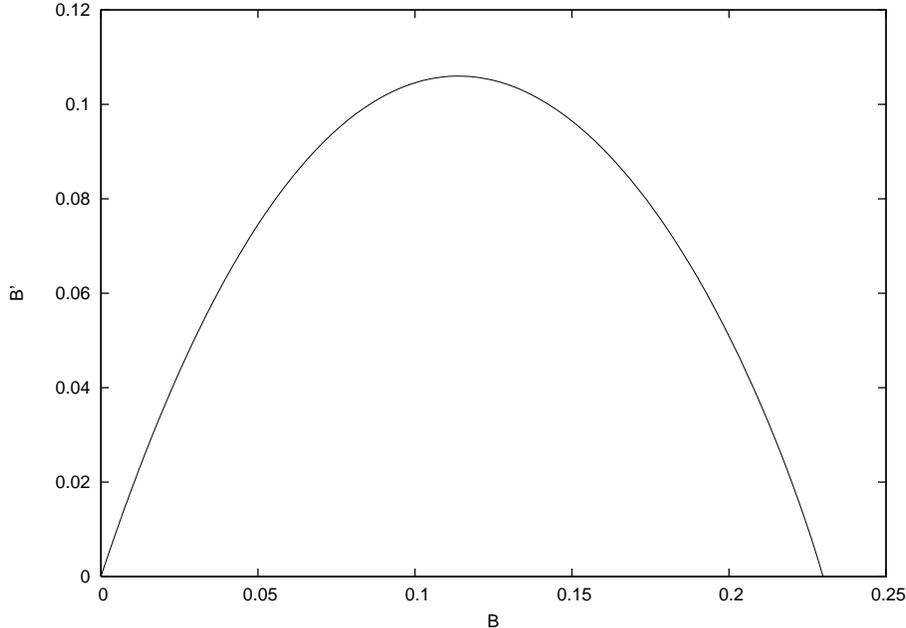}
\caption{\small{The orbit of the kink.  The node is approached along
the slow eigendirection $\mathbf{\xi}_1$.} }\label{fig6}
\end{figure}

%%%%%

\subsubsection{Analytic study of the kink solution}

In this Section we describe an attempt to find the kink solution in
the closed form. Although this attempt was not successful, we
believe it is worth presenting because it yields a deeper analytic
insight into the structure of equations. In addition, it provides an
alternative way of proving the existence of the kink.

   Consider the 8-metric
%%%%%
\be
ds_8^2= dr^2 + a^2 (\sigma_i-A^i)^2 + b^2 d\Omega_4^2\,,\label{abans}
\ee
%%%%%
where $d\Omega_4^2$ is the metric on the unit 4-sphere, $A^i$ is the
1-instanton solution on the $S^4$, and $\sigma_i$ are the left-invariant
1-forms of $SU(2)$.  The flat metric on $\R^8$ corresponds to $a=b=\ft12 r$,
in which case the principal orbits are round 7-spheres for all values
of $r$.  In the kink solution, which is easily found by numerically
solving the Ricci-flat equations for the metric ansatz (\ref{abans}),
the metric approaches the flat form at small $r$, whilst as $r$ goes
to infinity the principal orbits approach the squashed Einstein metric
on $S^7$, for which $a^2/b^2=1/5$.  In fact, as $r$ approaches
infinity the metric functions have the limiting forms
%%%%%
\be
a\longrightarrow \fft{3}{10}\, r\,,\qquad b\longrightarrow \fft{3}{2\sqrt5}\,
   r\,.
\ee
%%%%%
The nature of the numerical results can be seen in Figure 7 below:

\begin{figure}[h!bp]
\centerline{\epsfxsize=2.4truein
\epsffile{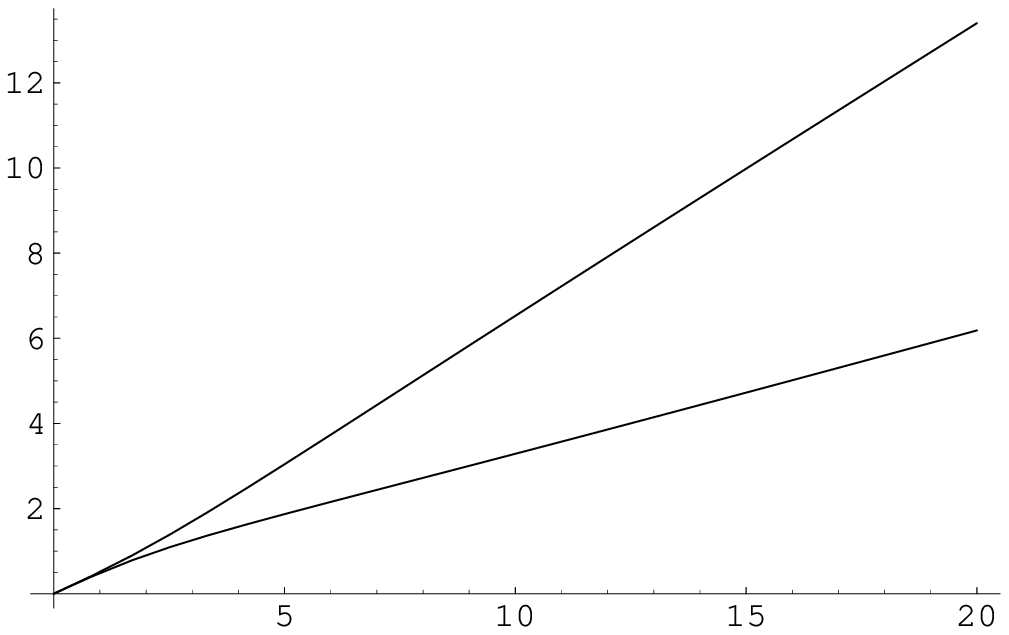}
\hspace{0.25in}
\epsfxsize=2.4truein
\epsffile{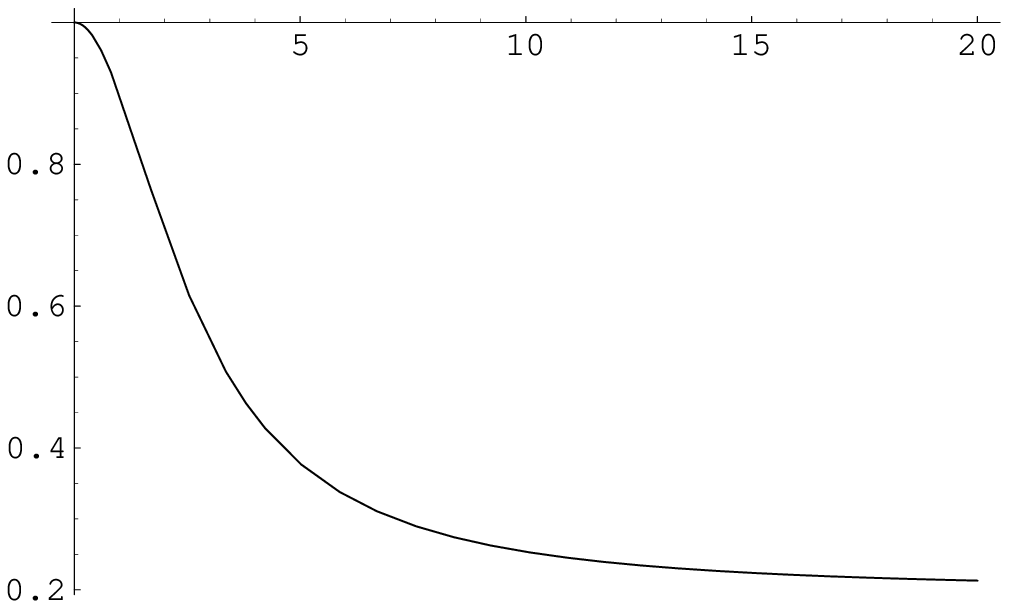}
}
\caption{The left-hand plot shows the functions $a$ (lower) and $b$ (upper)
for the kink solution written using the metric ansatz (\ref{abans}).
The right-hand plot
shows $a^2/b^2$, which ranges from 1 (the round $S^7$) at small $r$ to
$1/5$ (the squashed Einstein $S^7$) as $r$ goes to infinity.}
\end{figure}

  One can also attempt to solve analytically for the kink solutions.  (Note
that, up to scaling, there is a unique such solution in each dimension
$D=4n+4$.)  Let us again consider the eight-dimensional case, and change
variables so that (\ref{abans}) becomes
%%%%%
\be
ds_8^2 =\fft{e^{2\rho}\,  d\rho^2}{h(\rho)} +
        e^{2\rho}\, f(\rho)\, (\sigma_i - A^i)^2 + e^{2\rho}\,  d\Omega_4^2\,.
\label{kink8}
\ee
%%%%%
The specific choice of coordinate gauge in
(\ref{kink8}) is one that often allows one to obtain explicit solutions
in terms of rational functions of $e^\rho$.

   In the present case, we find that the metric is Ricci-flat if
%%%%%
\be
h= \fft{(1 + 8 f - 2f^2) f}{(8f^2 + 12 f \dot f +{\dot f}^2)}
\ee
%%%%%
and the function $f(\rho)$ satisfies the equation
%%%%%
\bea
&&2f(2f^2-8f-1)\ddot f +3(f-2){\dot f}^3 +2(21f^2-34f+2) {\dot f}^2\nn\\
&& +
  12f(17f^2-26f+2) \dot f + 56f^2(f-1)(5f-1)=0
\,,\label{feqn}
\eea
%%%%%
where a dot denotes a derivative with respect to $\rho$.
This equation can be reduced to a single first-order differential equation
as follows.  We define $f(\rho)=x$ and $\dot f(\rho)=y(x)$, which implies
%%%%%
\be
\rho= \int^x \fft{dz}{y(z)}\,,
\ee
%%%%%
and (\ref{feqn}) becomes
%%%%%
\bea
y' = \fft{3(x-2) y^3 + 2(2x^2-34x+2)y^2 + 12x(17x^2-26x+2) y
  + 56x^2(x-1)(5x-1)}{2 x(1+8x -2x^2)y}\,,\label{yeqn}
\eea
%%%%%
where the prime denotes a derivative with respect to $x$.  Note that
$x$ is the squashing parameter, ranging from $x=1$ near the origin
of the kink, where the 7-spheres are round, to $x=1/5$ in the
asymptotic region near infinity, where the 7-spheres approach the
squashed Einstein metric.  The function $h$ appearing in the metric
(\ref{kink8}) is given by
%%%%%
\be
h= \fft{x(1+8x-2x^2)}{8x^2 + 12 xy + y^2}\,.
\ee
%%%%%

   The kink solution corresponds to $y(x)$ in (\ref{yeqn}) describing
an arc, lying below the $y$ axis, starting at $(x,y)=(1/5,0)$ and
ending at $(x,y)=(1,0)$.  The asymptotic forms at the two endpoints of
the arc are
%%%%%
\bea
y&=& -\fft8{15} (5x-1) - \fft{26}{105} (5x-1)^2 + \fft{1954}{5145}
    (5x-1)^3 +\cdots\,,\nn\\
y&=& -2(1-x) + \fft{24}{7} (1-x)^2 - \fft{640}{343} (1-x)^3 +\cdots
\eea
%%%%%
respectively.
In terms of the original variables, the kink solution runs from
%%%%%%
\be
f\sim 1\,,\qquad h\sim \fft14
\ee
%%%%%
near the origin at $\rho=-\infty $ to
%%%%%
\be
f\sim \fft{1}{5}\,,\qquad h\sim \fft{9}{20}
\ee
%%%%%
as $\rho$ approaches infinity.

   It should be remarked that there exists a known solution to
the equations, corresponding to the complete metric of Spin(7) holonomy
found in \cite{brysal,gibpagpop}.  This corresponds to
%%%%%
\be
f(\rho) = \fft15  (1- e^{-10\rho/3})\,,\qquad \hbox{or}\qquad
  y(x) = \fft23 (1-5x)\,.\label{spin7sol}
\ee
%%%%%
This solution has the same behaviour at large $\rho$ (\ie $x\rightarrow 1/5$)
as we require for the kink solution, but it is very different at short
distance (corresponding to $\rho=0$ in this parameterisation), since it
has an $S^4$ bolt.  In fact in the Spin(7) solution the variable $x$
lies in the range $0\le x\le 1/5$.

   By making further transformations, one can cast (\ref{yeqn}) into
a standard form for an Abel equation of the first kind.  First, we
make use of the
known solution given in (\ref{spin7sol}), and define
a new dependent variable $v(x)$, related to $y(x)$ by
%%%%%
\be
y(x) = \Big(\fft{2x^2-8x-1}{x^2(5x-1)v(x)} - \fft{3}{2(5x-1)}\Big)^{-1}\,.
\ee
%%%%%
After this change of variable, equation (\ref{yeqn}) becomes
%%%%%
\be
x^3(5x-1) v v' = x^3 v^2 + 6x(4x^2+5x-1) v + 28(1-x)(2x^2-8x-1)\,.
\label{veqn}
\ee
%%%%%
The further change of variable to $u(x)=(5x-1)^{-1/5} v(x)$ yields
%%%%%
\be
x^3(5x-1)^{7/5}\, u u' = 6x(5x-1)^{1/5} (4x^2+5x-2) u + 28(1-x)(2x^2-8x-1)\,.
\label{ueqn}
\ee
%%%%%
It seems not to be possible to carry this further, since the change
of independent variable required to put the equation into the canonical
form $u du/dz -u = q(z)$ is given by taking
%%%%%
\be
z= \fft{6(31x-2)}{x(5x-1)^{1/5}} - 42 \,\,5^{4/5} x^{-1/5}\, _
  2F_1(\ft15, \ft15; \ft65; \ft1{5x})\,,\label{zx}
\ee
%%%%%
yielding
%%%%%
\be
u \fft{du}{dz} - u = \fft{14(1-x)(2x^2-8x-1)}{3 x(5x-1)^{1/5}(4x^2+5x-1)}\,.
\ee
%%%%%
Since one cannot invert (\ref{zx}) explicitly to obtain $x$ as a function
of $u$, it appears that no further progress towards an analytic solution
can be made.

   Although attempts to solve (\ref{yeqn}) completely by analytic means
have not proved successful, one can use this first-order system to
perform a phase-plane analysis.  This can be done
by writing the first-order
equation (\ref{yeqn}) in terms of an auxiliary parameter $t$, with
%%%%%
\bea
\fft{dx}{dt} &=& 2 x(1+8x -2x^2)y\,,\nn\\
\fft{dy}{dt} &=& 3(x-2) y^3 + 2(2x^2-34x+2)y^2 + 12x(17x^2-26x+2) y
  \nn\\
&&+ 56x^2(x-1)(5x-1)\,.
\eea
%%%%%
The kink solution lies within the region $1/5\le x\le 1$ in the
phase plane. In fact it corresponds to the unique flow that starts
at the saddle at $(x,y)=(1,0)$, and ends at the attractor at
$(x,y)=(1/5,0)$. It is also instructive to look at the exact Spin(7)
solution (\ref{spin7sol}). This starts at the saddle at
$(x,y)=(0,2/3)$, and flows (along a straight line) to the attractor
at $(x,y)=(1/5,0)$.  The kink solution, and the extrapolation of the
exact Spin(7) solution into the region $1/5\le x\le 1$, are shown in
Figure 8 below.

\begin{figure}[h!bp]
\centerline{\epsfxsize=2.4truein
\epsffile{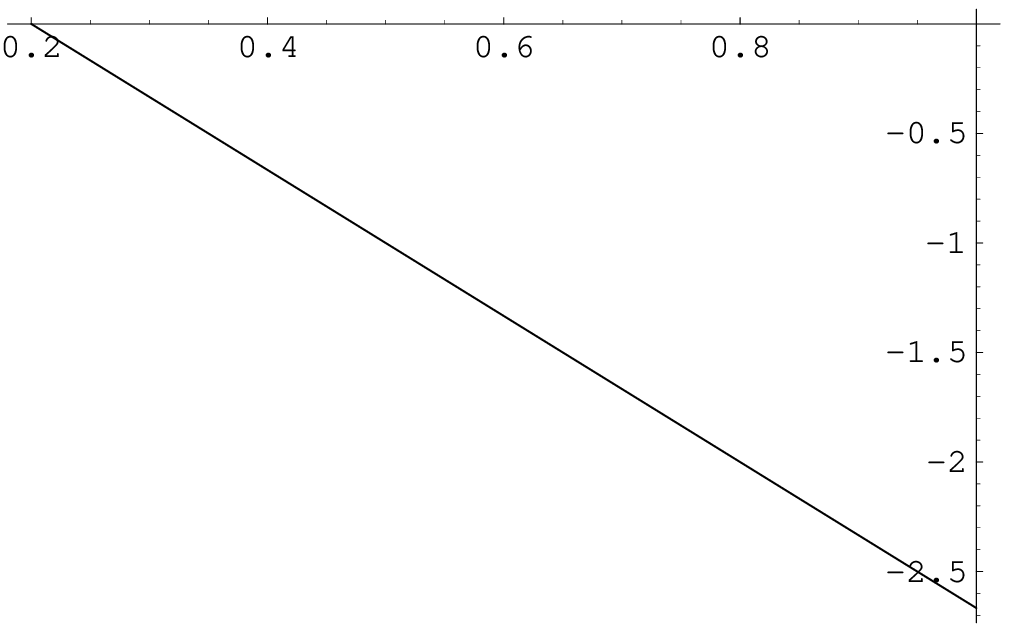}
\hspace{0.25in}
\epsfxsize=2.4truein
\epsffile{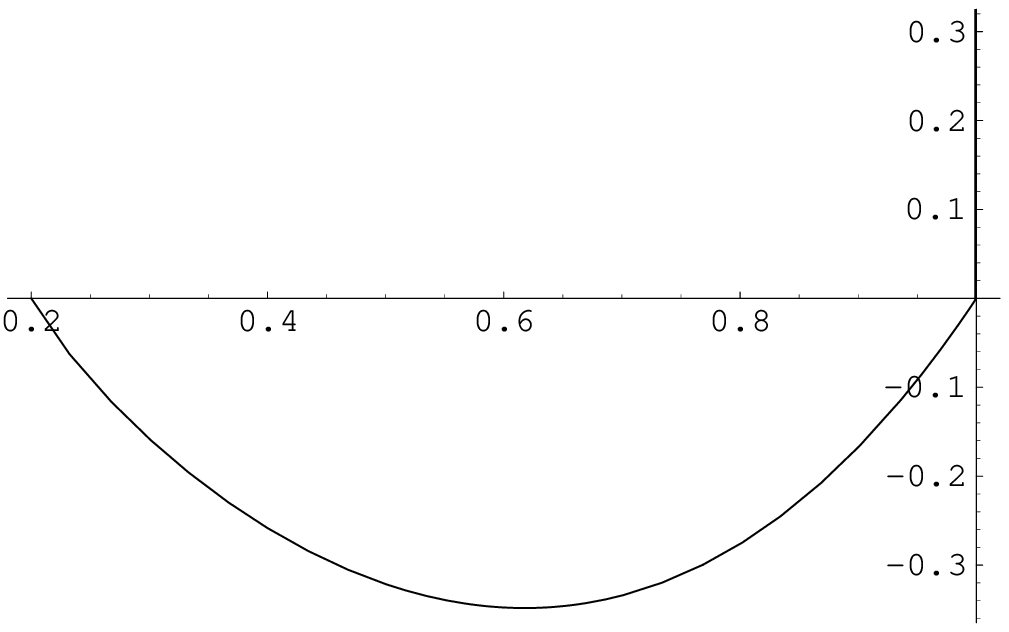}
}
\caption{The numerical solutions of (\ref{yeqn}) corresponding,
in the left-hand plot, to the exact solution of Spin(7) holonomy given in
(\ref{spin7sol}) (extrapolated into the region $1/5\le x\le 1$ depicted here),
and in the right-hand diagram, to the kink solution.
The latter corresponds to a trajectory from the saddle at $x=1$, $y=0$
to the attractor at $x=1/5$, $y=0$.}
\end{figure}

\subsubsection{Stability of the kink soliton}

In the case of the kink  the eigenvalue problem (\ref{stabcont}) for
the spectrum of small perturbations around a static solution reduces
to the single equation  \be -\frac{1}{r^7}
e^{\delta_0}\Big(e^{\delta_0} r^7 v'_{\lambda}\Big)'+ K v_{\lambda}
= \lambda^2 v_{\lambda}\,, \ee
 where \bea
 K&=& \frac{1}{7 r^2} \Big(-72
e^{-3B_0}-16 e^{4 B_0} + 200 e^{-200 B_0} \Big) \\
\nonumber &+&\frac{1}{7 r} \Big(288 e^{-3B_0}-48 e^{4 B_0} - 240
e^{-200 B_0} \Big) B'_0 \\ \nonumber &+& \frac{1}{7} \Big(-288
e^{-3B_0}-36 e^{4 B_0} + 72 e^{-200 B_0} \Big) {B'_0}^2\,. \eea
We have shown above that the profile function of the kink is
monotone, hence the zero mode corresponding to the scaling freedom,
$v_0=r B_0'(r)$, has no zeros which implies by the standard
Sturm-Liouville theory there are no negative eigenvalues. Thus, the
kink solution is linearly stable within our ansatz.

The behavior of the kink under non-infinitesimal perturbations was
studied numerically by the methods described in Section 3.4.2. As in
the case of other static solutions we found that for small
perturbations the kink is asymptotically stable, while  for large
perturbations it collapses to a black hole. The numerical evidence
for these properties is shown in Figs.~9 and~10.
\begin{figure}[h!]
\centering \subfigure[\normalsize{}]{ \label{fig:subfig:a}
\includegraphics[width=0.7\textwidth]{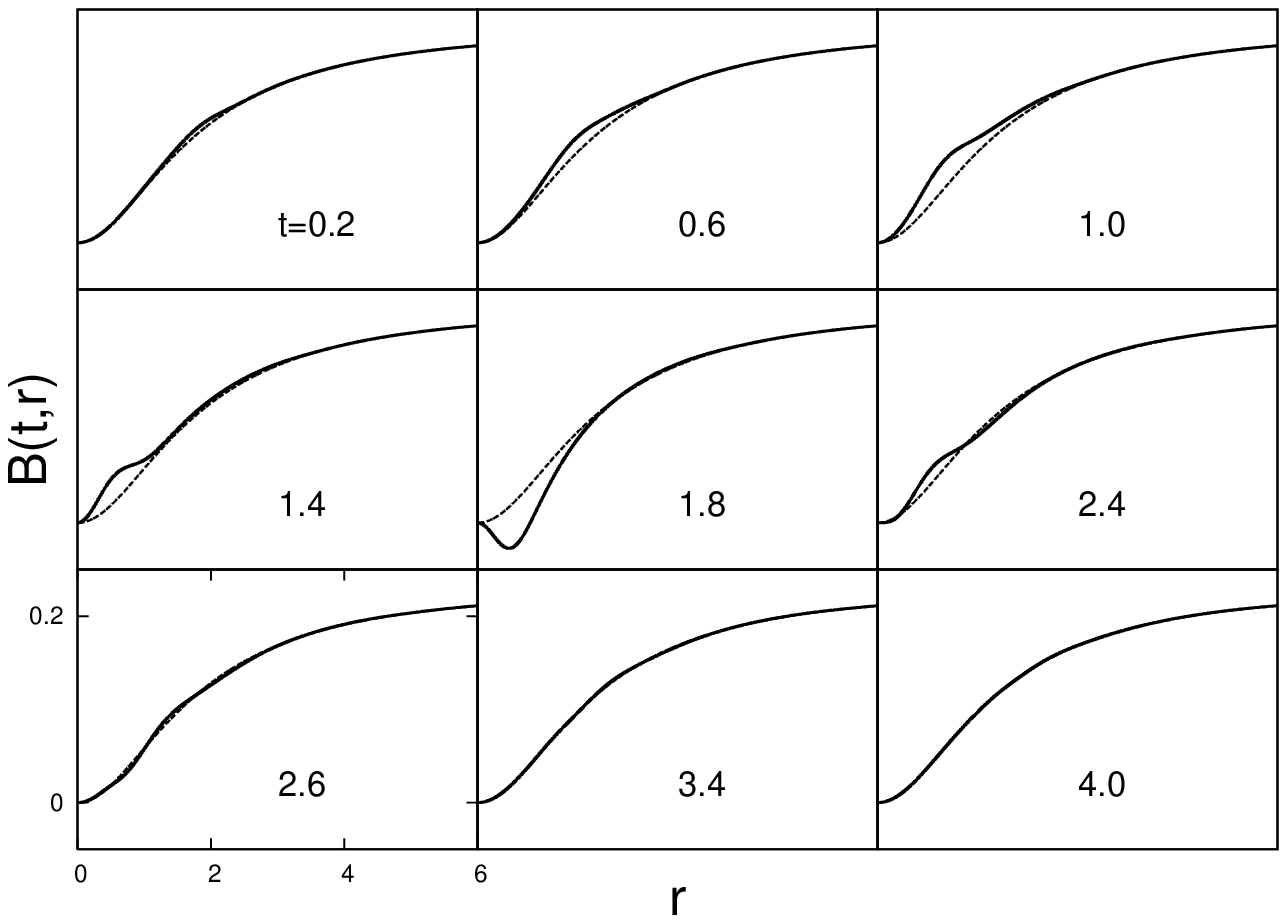}}
\subfigure[\normalsize{}]{ \label{fig:subfig:b}
\includegraphics[width=0.7\textwidth]{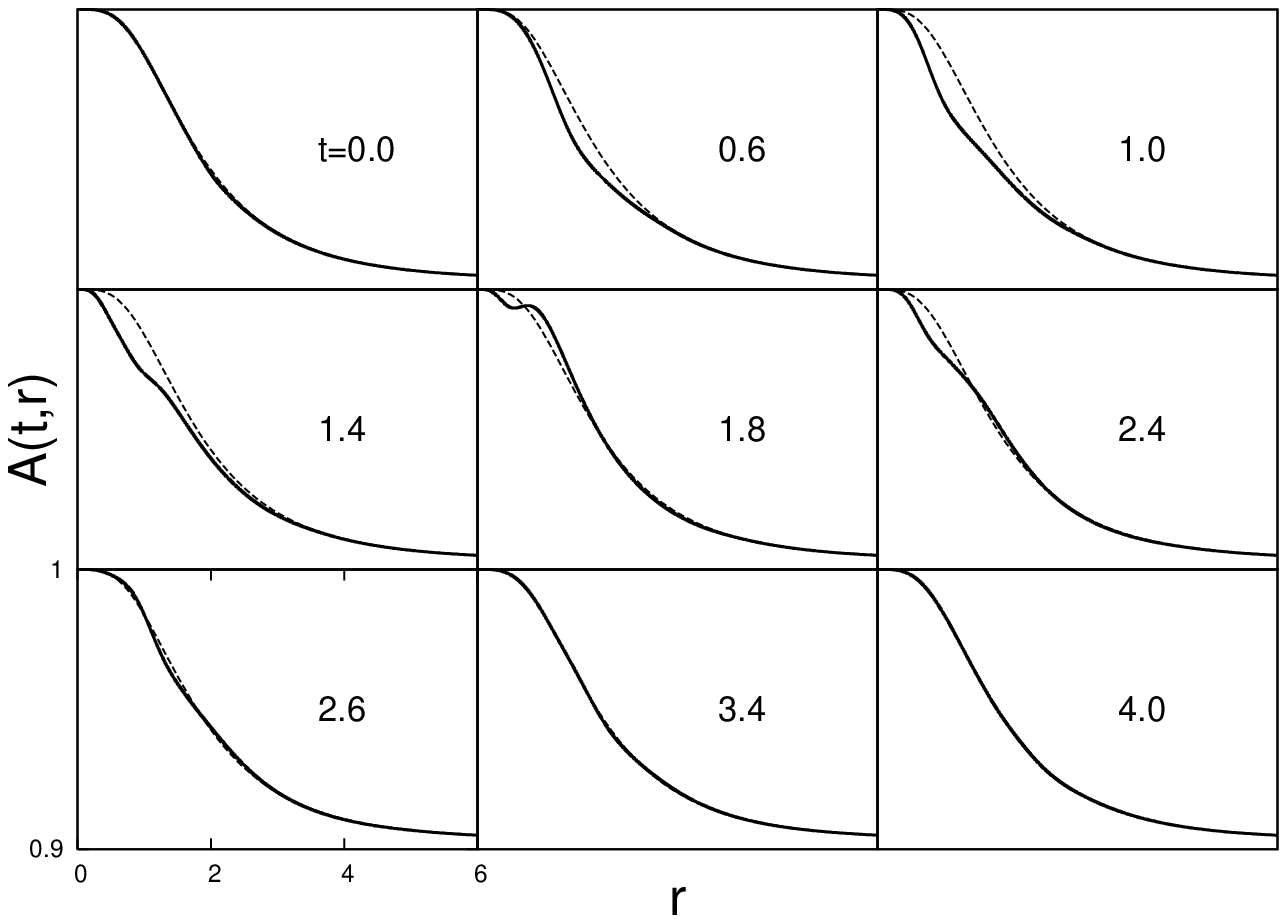}}
\caption{\small{Asymptotic stability of the kink solution. For small
perturbations of the kink we plot a series of snapshots of (a) the
function $B(t,r)$ and (b) the function $A(t,r)$. The dashed line
shows the unperturbed kink solution.}} \label{fig:subfig}
\end{figure}
\newpage

\begin{figure}[h!]
\centering \subfigure[\normalsize{}]{ \label{fig:subfig:a2}
\includegraphics[width=0.65\textwidth]{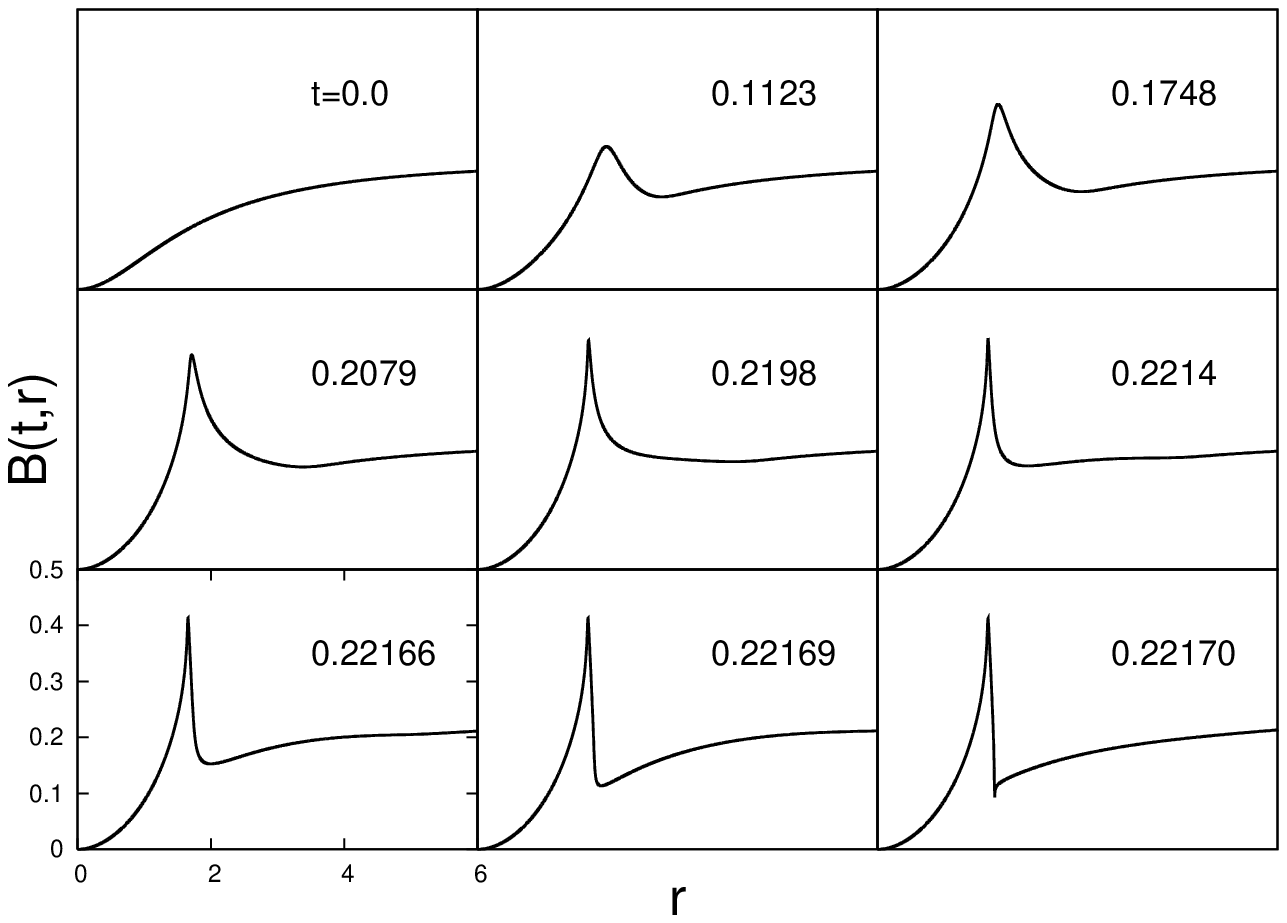}}
\subfigure[\normalsize{}]{ \label{fig:subfig:b2}
\includegraphics[width=0.65\textwidth]{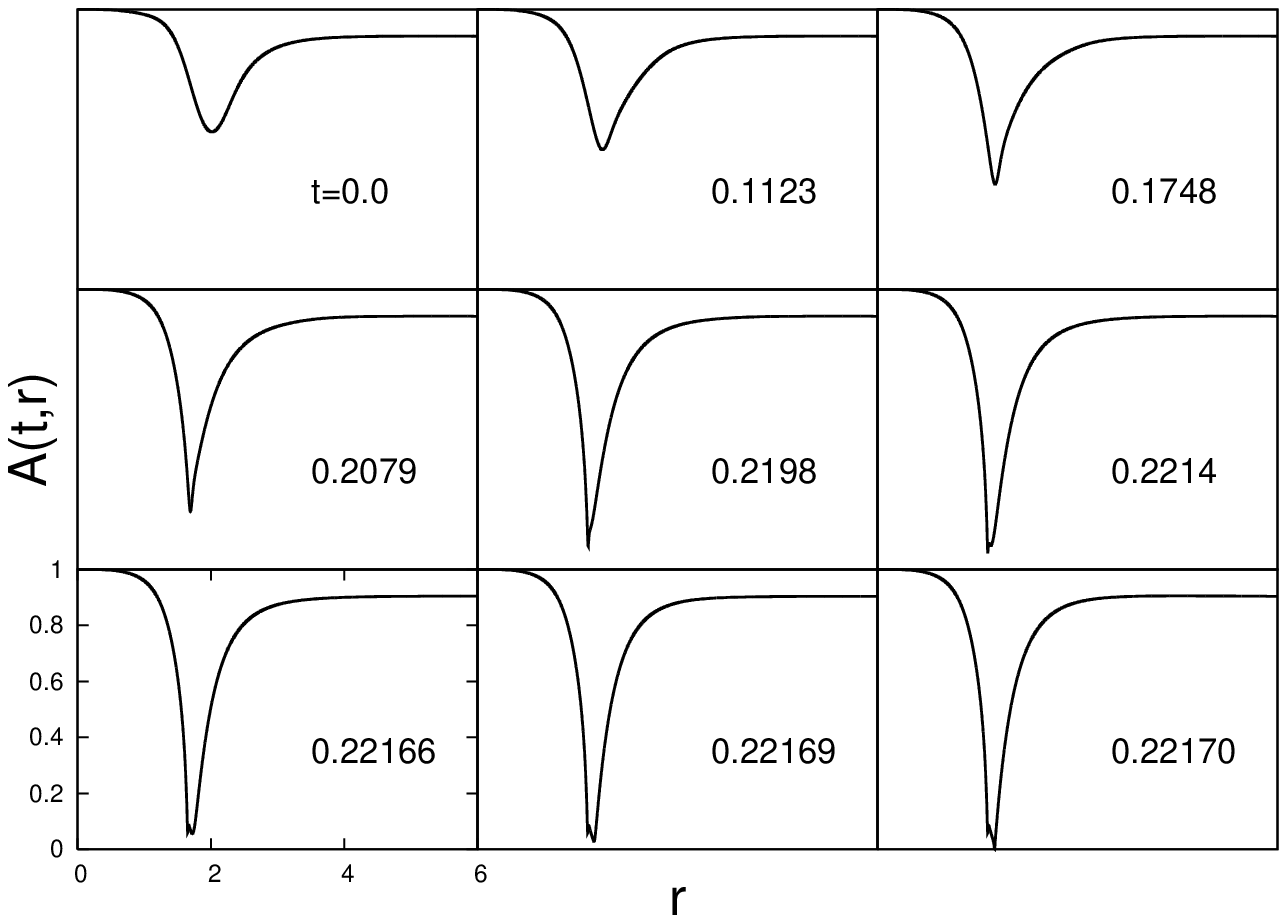}}
\caption{\small{Instability of the kink for large perturbations. For
large perturbations of the kink we plot a series of snapshots of (a)
the function $B(t,r)$ and (b) the function $A(t,r)$. During the
evolution $A(t,r)$ drops to zero at $r=r_H \approx 1.754$ which
signals the formation of an apparent horizon there. Outside the
horizon the solution relaxes to a static black hole.}}
\label{fig:subfig2}
\end{figure}

\section{Conclusions}

   In this paper, we have studied gravitational solitons and kinks in
higher dimensions.  Our focus has been the study of their
stability, principally in the case of solitons in nine dimensions.  We
have considered various possibilities for the spatial metric,
including examples such as the higher-dimensional
Taub-NUT metrics \cite{baibat,pagpop}, which
are not supersymmetric, and also the example of the $\A8$ 8-metric of
Spin(7) holonomy \cite{spin70,spin7}, which is supersymmetric.  All the
solitons we consider are trivial topologically (\ie $\R^n$ topology), but
non-trivial geometrically.  We studied the question of stability first
at the linearised level, using analyic methods, and found that in all cases
the solitons are linearly stable.  Numerical analysis indicates that this
stability persists for non-infinitesimal perturbations also, provided
they are small enough in magnitude.  The numerical analysis also
shows that for sufficiently large perturbations the solitons are all
unstable to black-hole formation.  This is not altogether surprising,
since even flat Minkowski
spacetime is unstable to sufficiently large perturbations, which can lead
to the formation of a black hole.   These instabilities provide a salutory
reminder of the fact that in gravitational theories supersymmetry is
not a guarantor of stability beyond the linearised level.

   We also studied in detail a kink metric in eight dimensions, whose
existence was first encountered in \cite{pagpop2}.  It is a non-trivial
cohomogeneity-1 metric on $\R^8$, in the which the level sets are
homogeneous 7-spheres viewed as $S^3$ bundles over $S^4$.  At small
distances the level surfaces approach the round 7-sphere and the metric
is of the form near the origin of Euclidean 8-space.  At large
distances the level surfaces approach the squashed $S^7$ Einstein metric
in the family of $S^3$ bundles over $S^4$.   We showed that the Ricci-flatness
conditions for the kink metric can be reduced to a first-order Abel equation
of the first kind, but it appears not to be possible to obtain an
explicit solution analytically.  Our discussion includes a proof of the
existence of the Ricci-flat metric.

\subsection*{Acknowledgments.} The research of PB and TC was
supported in part by the Polish Research Committee grant 1PO3B01229.
The research of C.N.P. is supported in part by DOE grant DE-FG03-95ER40917.

\appendix

\section{Nine-Dimensional Schwarzschild}

   This section contains a brief summary of the analysis of the stability
of the nine-dimensional Schwarzschild solution within the framework of
the deformations considered in this paper.

   If we expand around the nine-dimensional Schwarzschild background,
for which
%%%%%
\be A_0= 1-\fft{2m}{r^6}\,,\qquad \delta_0=0\,,\qquad
\phi_1^{(0)}=0\,,\qquad \phi_2^{(0)}=0\,, \ee
%%%%%
we find that working to first order in fluctuations we can keep
$A=A_0$ and $\delta=0$.   For the dynamical fields, we shall now use
$\phi_1$ and $\phi_2$ to denote the linearised fluctuations.
Introducing the ``tortoise coordinate'' $x$ via $dx/dr= A_0^{-1}$,
and defining
%%%%%%
\be \phi_1 = r^{-7/2}\, u_1\,,\qquad \phi_2= r^{-7/2}\, u_2\,, \ee
%%%%%%
we find that these satisfy
%%%%%
\be \ddot u_1 - \fft{\del^2 u_1}{\del x^2} + V\, u_1 =0\,,\qquad
\ddot u_2 - \fft{\del^2 u_2}{\del x^2} + V\, u_2 =0\,, \ee
%%%%%
where the potential $V$ (which is the same for both $u_1$ and $u_2$)
is given by
%%%%
\be V= -\ft14  \Big(1- \fft{2m}{r^6}\Big)
  \Big(\fft{99}{r^2} + \fft{98m}{r^8}\Big)\,.
\ee
%%%%%
Note that this is the same potential as was encountered in
\cite{BCS2} in the analysis of the nine-dimensional perturbations of
Schwarzschild with a single dynamical variable (corresponding to
$u_2=0$ here.)

\section{Time Evolution of Kerr-AdS Black Holes}

   In this appendix, we use the techniques of this paper to analyse
the stability of a particular class of five-dimensional rotating
black holes.  The general methods extend extend to any black hole
having all angular momenta equal, but in this appendix we restrict
attention to the $(4+1)$-dimensional case.  This means that we can
again consider an ansatz
with $SU(2)\times U(1)$ isometry on the constant-radius spatial
sections.

   The five-dimensional Kerr-AdS solution \cite{hawhuntay}
with equal angular momenta is given by
%%%%%
\be
ds_5^2 = - \fft{G dt^2}{\Big(1 + \fft{2 M a^2}{\rho^4}\Big)}
  + \fft{d\rho^2}{G} + \ft14 \rho^2\Big(1+ \fft{2M a^2}{\rho^4}\Big)
(\sigma_3 + 2\Omega\, dt)^2
 + \ft14 \rho^2\, (\sigma_1^2 + \sigma_2^2)\,,\label{kerrads}
\ee
%%%%%
where
%%%%%
\bea
G &=&\Big( 1 - \lambda\, \rho^2 - \fft{2M\, \Xi}{\rho^2} +
   \fft{2M a^2}{\rho^4}\Big)\,,\nn\\
\Omega &=& \fft{2M a}{\rho^4 + 2 M a^2}\,,
\qquad \Xi= 1 +\lambda a^2\,.\label{kerrsol}
\eea
%%%%%
The metric satisfies $R_{MN}= 4 \lambda g_{MN}$.  Thus for the asymptotically
AdS case we require $\lambda<0$.

   The new feature in our analysis is the inclusion of the Kaluza-Klein
vector in the dimensional reduction.  The electric charge associated with
this field is proportional to the angular momentum of the black hole.
Thus we make the following reduction ansatz
%%%%%
\be
d\hat s^2 = e^{2\alpha\,\varphi} \, ds^2 + e^{2\beta\, \varphi}\,
\Big[ e^{4\gamma\, \phi}\, (\sigma_3 + {\cal A})^2 + e^{-2\gamma\, \phi}\,
   (\sigma_1^2 + \sigma_2^2)\Big]\,.\label{metred}
\ee
%%%%%
For the time being, we consider the base metric $ds^2$ to have dimension
$n$.  Later,
we shall specialise to the case of immediate interest, namely $n=2$.
We choose the natural vielbein basis
%%%%%
\bea
\hat e^0 &=& e^{\beta \varphi + 2\gamma \phi}\, (\sigma_3 + {\cal A})\,,\nn\\
\hat e^\alpha &=& e^{\alpha \varphi}\, e^\alpha\,,\nn\\
\hat e^a &=& e^{\beta \varphi -\gamma \phi}\, \sigma_a\,,\qquad
   a=1,2\,.
\eea
%%%%%

   After some calculation, we arrive at the following non-vanishing
Ricci-tensor components (in the vielbein basis):
%%%%%
\crampest
\bea
\hat R_{00} &=& e^{-2\alpha \varphi}\,
\Big\{ -(3\beta + (n-2)\alpha) [\beta(\nabla\varphi)^2
  + 2\gamma\nabla\varphi\cdot\nabla\phi] -\beta\square\varphi - 2\gamma
     \square\phi \Big\} \nn\\
  && + \ft14 e^{(2\beta-4\alpha)\varphi + 4\gamma\phi}\, {\cal F}^2 +
    \ft12 e^{-2\beta\varphi + 8\gamma\phi}\,,\nn\\
\hat R_{\alpha\beta} &=& e^{-2\alpha\varphi}\, \Big\{
(6\alpha\beta -3\beta^2+(n-2)\alpha^2)\nabla_\alpha\varphi\,
\nabla_\beta\varphi - 6\gamma^2\nabla_\alpha\phi\, \nabla_\beta\phi
 - (3\beta+(n-2)\alpha)\nabla_\alpha\nabla_\beta\varphi\Big\}\nn\\
&& - \alpha e^{-2\alpha\varphi}\,
\Big\{(3\beta+(n-2)\alpha)(\nabla\varphi)^2
  + \square\varphi\Big\}\eta_{\alpha\beta} - \ft12 e^{(2\beta-4\alpha)\varphi
  + 4\gamma\phi}\, {\cal F}_{\alpha\gamma}\, {\cal F}_{\beta}{}^\gamma +
    e^{-2\alpha\varphi}\, R_{\alpha\beta}\,,\nn\\
\hat R_{ab} &=& \Big[ e^{-2\alpha\varphi}\, \Big\{
   (3\beta + (n-2)\alpha)[\gamma \nabla\varphi\cdot\nabla\phi - \beta
  (\nabla\varphi)^2] + \gamma\square\phi-\beta\square\varphi\Big\}\nn\\
  &&\qquad - \ft12 e^{-2\beta\varphi + 2\gamma\phi}\, (e^{6\gamma\phi}-2)\Big]
\, \delta_{ab}\,,\nn\\
\hat R_{0\alpha} &=&
\ft12 e^{(\alpha-\beta)\varphi -2\gamma\phi}\, \nabla_\beta\Big(
  e^{(2\beta-4\alpha)\varphi + 4\gamma\phi}\, {\cal F}_\alpha{}^\beta\Big)
  + \ft12 (n\alpha+ 3\beta)\, e^{(\beta-3\alpha)\varphi + 2\gamma\phi}\,
    {\cal F}_\alpha{}^\beta\, \nabla_\beta\varphi\,,
\uncramp
\eea
%%%%%
where ${\cal F}=d{\cal A}$.

   Specialising to the case $n=2$, these expressions give
%%%%%
\bea
\hat R_{00} &=& e^{-2\alpha \varphi}\,
\Big\{ -3\beta [\beta(\nabla\varphi)^2
  + 2\gamma\nabla\varphi\cdot\nabla\phi] -\beta\square\varphi - 2\gamma
     \square\phi \Big\} \nn\\
  && + \ft14 e^{(2\beta-4\alpha)\varphi + 4\gamma\phi}\, {\cal F}^2 +
    \ft12 e^{-2\beta\varphi + 8\gamma\phi}\,,\label{R00}\\
\hat R_{\alpha\beta} &=& e^{-2\alpha\varphi}\, \Big\{
3\beta(2\alpha -\beta)\nabla_\alpha\varphi\,
\nabla_\beta\varphi - 6\gamma^2\nabla_\alpha\phi\, \nabla_\beta\phi
 - 3\beta\nabla_\alpha\nabla_\beta\varphi\Big\}\nn\\
&& - \alpha e^{-2\alpha\varphi}\,
\Big\{3\beta (\nabla\varphi)^2
  + \square\varphi\Big\}\eta_{\alpha\beta} - \ft12 e^{(2\beta-4\alpha)\varphi
  + 4\gamma\phi}\, {\cal F}_{\alpha\gamma}\, {\cal F}_{\beta}{}^\gamma +
    e^{-2\alpha\varphi}\, R_{\alpha\beta}\,,\label{Ralbe}\\
\hat R_{ab} &=& \Big[ e^{-2\alpha\varphi}\, \Big\{
   3\beta[\gamma \nabla\varphi\cdot\nabla\phi - \beta
  (\nabla\varphi)^2] + \gamma\square\phi-\beta\square\varphi\Big\}
  - \ft12 e^{-2\beta\varphi + 2\gamma\phi}\, (e^{6\gamma\phi}-2)\Big]
\, \delta_{ab}\,,\label{Rab}\\
\hat R_{0\alpha} &=&
\ft12 e^{-(\alpha+4\beta)\varphi -2\gamma\phi}\, \nabla_\beta\Big(
  e^{(5\beta-2\alpha)\varphi + 4\gamma\phi}\, {\cal F}_\alpha{}^\beta\Big)\,.
\label{2deqns}
\uncramp
\eea
%%%%%

   Following the analysis in \cite{BCS}, we use the volume of the
$S^3$ to parameterise the
radial direction, and write
%%%%%
\be
d\hat s^2 = -A e^{-2\delta}\, dt^2 + A^{-1}\, dr^2 +
  \ft14 r^2 [ e^{2B} (\sigma_1^2 + \sigma_2^2) + e^{-4B}\,
    (\sigma_3+ {\cal A})^2]\,.\label{metans}
\ee
%%%%%
Thus, comparing with (\ref{metred}), we have
%%%%%
\be
e^{\beta\varphi}= \ft12 r\,,\qquad \gamma\phi= -B\,,\qquad
ds^2 = e^{-2\alpha\varphi}\, (- A e^{-2\delta}\, dt^2 + A^{-1}\, dr^2)\,.
\ee
%%%%%
Substituting into the Einstein equation
%%%%%
\be
\hat R_{MN}= 4\lambda\, \hat g_{MN}\,,
\ee
%%%%%
we note from (\ref{2deqns}) that
%%%%%
\be
{\cal F}_{\mu\nu} =
c\, \ep_{\mu\nu} \, e^{(2\alpha-5\beta)\varphi - 4\gamma\phi}\,,
\label{Fsol}
\ee
%%%%%
where $c$ is a constant and $\ep_{\mu\nu}$ is the Levi-Civita tensor in
the metric $ds^2$.  This expression can be substituted into the remaining
Einstein equations, leading to the momentum and Hamiltonian constraints
%%%%%
\bea
A' &=& -\fft{2A}{r} + \fft1{3r}\, (8e^{-2B} - 2 e^{-8B}) -
  2r(e^{2\delta} A^{-1}\, \dot B^2 + A {B'}^2) - \fft{128 c^2}{3 r^7}\,
     e^{4B}- 4\lambda\, r\,,\nn\\
\dot A &=& - 4 r A \dot B B'\,,\label{constraints}
\eea
%%%%%
the slicing condition
%%%%%
\be
\delta' = -2r (e^{2\delta} \, A^{-2}\, \dot B^2 + {B'}^2)\,,\label{slice}
\ee
%%%%%
and the wave equation
%%%%%
\bea
\Big( e^\delta\, A^{-1}\, r^3\, \dot B\Big)^{.} -
 \Big(e^{-\delta}\, A\, r^3\, B'\Big)' + \ft43 e^{-\delta}\, r\,
  (e^{-2B} - e^{-8B}) + \fft{128 c^2}{3 r^5}\, e^{-\delta}\, e^{4B}=0\,.
\label{wave}
\eea
%%%%%

   It is straightforward to verify the self consistency of the
equations.  Namely, that the vanishing of the dot of $A'$ minus
the prime of $\dot A$ in (\ref{constraints}) yields, after using
(\ref{slice}) and (\ref{constraints}), the wave equation (\ref{wave})
for $B$.

   Defining $A= 1 - m(r,t)/r^2 - \lambda r^2$, the Hamiltonian
constraint in (\ref{constraints}) becomes
%%%%%
\be
m' = 2 r^3 (e^{2\delta}\, A^{-1}\, \dot B^2 + A {B'}^2) +
 \ft23 r( 3 + e^{-8B} - 4 e^{-2B}) + \fft{128 c^2}{3 r^5}\, e^{4B}\,.
\ee
%%%%%
This is manifestly positive.

   Note that the constant $c$ is related to the angular momentum.
Lowering the index on the Killing vector $\del/\del\psi$ gives the
1-form
%%%%%
\be
K= e^{2\beta\varphi + 4\gamma\phi}\, (\sigma_3 + {\cal A})\,.
\ee
%%%%%
The angular momentum is given by the Komar integral $J=1/(16\pi)\,
  \int_{S^3} {\hat *dK}$, where $\hat *$ is the Hodge dual in
the five-dimensional metric $d\hat s_5^2$, and we have
%%%%%
\be
{\hat *dK} = e^{(5\beta-2\alpha)\varphi + 4\gamma\phi}\, {*{\cal F}}\wedge
 \sigma_1\wedge\sigma_2\wedge\sigma_3 + \cdots\,,
\ee
%%%%%
where ${*{\cal F}}$ is the 0-form Hodge dual of the field strength
${\cal F}$
in the 2-metric $ds^2$.  Thus the
angular momentum can be seen to be nothing but the 2-dimensional
electric charge of the dimensionally-reduced solution.  From (\ref{Fsol}),
the Komar integral therefore gives the angular momentum
%%%%%
\be
J= \fft{c}{16\pi}\, \int_{S^3} \sigma_1\wedge\sigma_2\wedge\sigma_3
= \pi\, c\,.
\ee
%%%%%

Comparing the radial coordinate $r$ used in (\ref{metans}) with
the radial coordinate $\rho$ used in (\ref{kerrads}), we see that
%%%%%
\be
r = \rho\, \Big(1+ \fft{2M a^2}{\rho^4}\Big)^{1/6}\,.
\ee
%%%%%
This can be used to rewrite the previous equations using
$\rho$ instead of $r$ as the radial variable.  We make the expansion
%%%%%
\be
A= A_0 (1+ \wtd A)\,,\qquad B= B_0 + \wtd B\,,\qquad
 \delta= \delta_0 + \td\delta\,,
\ee
%%%%
where $A_0$, $B_0$ and $\delta_0$ denote the background expressions in
the Kerr-AdS metric, which can be read off from (\ref{kerrads}) and
(\ref{kerrsol}), and we work to linear order in the tilded quantities.

 From the slicing equation (\ref{slice}) we obtain
%%%%%
\be
(3 \rho^4 + 2M a^2)\, \fft{\del\td\delta}{\del\rho} +
   16 M a^2\, \fft{\del\wtd B}{\del\rho}=0\,,
\ee
%%%%%
whilst from the Hamiltonian constraint in (\ref{constraints}) we find
%%%%%
\be
(3\rho^4 + 2 M a^2) \wtd A + 16 M a^2 \wtd B=0\,.\label{ham2}
\ee
%%%%%
(Taking a constant of integration to be zero.)
These two equations can be used to solve for the perturbations
$\td\delta$ and $\wtd A$ in terms of the dynamical variable
$\wtd B$.\footnote{Note that in the case of linearisation around
the Schwarzschild solution, discussed in \cite{BCS}, one can
take $\td\delta=0$ and $\wtd A=0$, so that only the perturbation
$\wtd B$ of the dynamical variable $B$ need be considered in that case.}
The momentum constraint in (\ref{constraints}) implies
%%%%%
\be
(3\rho^4 + 2 M a^2) \dot{\wtd A} + 16 M a^2 \dot{\wtd B}=0\,,
\ee
%%%%%
which is consistent with (\ref{ham2}).

    The wave equation (\ref{wave}) then gives
%%%%%
\bea
\fft{\rho^3}{G}\Big(1 + \fft{2Ma^2}{\rho^4}\Big) \ddot {\wtd B}
   - (\rho^3\, G\, \wtd B')' +
\fft{8\rho(3 + \fft{12M a^2}{\rho^4} +
     \fft{16 M^2 a^2(3+ 2 \lambda a^2)}{3\rho^6} -
 \fft{4 M^2 a^4}{\rho^8}\Big)}{3\Big(1 + \fft{2M a^2}{3\rho^4}\Big)^2}
 \wtd B=0\,,\label{wave2}
\eea
%%%%%
where a prime here means $d/d\rho$, and, from (\ref{kerrsol},
%%%%%
\be
G\equiv 1 - \lambda \rho^2 - \fft{2M\Xi}{\rho^2} +
\fft{2M a^2}{\rho^4}\,.\label{Gdef}
\ee
%%%%%

   We can cast the wave equation into a Schr\"odinger form by introducing
the ``tortoise coordinate'' $x$ defined by
%%%%%
\be
dx = \Big(1 + \fft{2M a^2}{\rho^4}\Big)^{1/2}\, G^{-1}\, d\rho\,,
\ee
%%%%%
and introducing a new
dynamical variable $\chi(x,t)$, defined by
%%%%%
\be
\wtd B = \rho^{-3/2}\, \Big(1 + \fft{2M a^2}{\rho^4}\Big)^{-1/4}\, \chi\,.
\ee
%%%%%
Equation (\ref{wave2}) then takes the form
%%%%%
\be
\ddot\chi - \fft{\del^2\chi}{\del x^2} + V\, \chi=0\,,
\ee
%%%%%
where the potential $V$ is given by
%%%%%
\bea
V&=& \fft{G}{4\rho^2}\, \Big[ 35 +15\lambda \rho^2 +
  \fft{160 M^3 a^4(1+2\lambda a^2) \rho^2}{(\rho^4+ 2 M a^2)^3} -
  \fft{16M^2 a^2(7\rho^2+ 15\lambda a^2 \rho^2 +a^2)}{(\rho^4+ 2 M a^2)^2}\nn\\
&& \qquad-
   \fft{2M(87 \rho^2 + 58\lambda a^2 \rho^2 -92 a^2)}{\rho^4+ 2 M a^2} -
 \fft{256 M^2 a^2(3\rho^2 +2\lambda a^2\rho^2 -2 a^2)}{(3\rho^2 + 2M a^2)^2}
\nn\\
&& \qquad+
 \fft{64 M(9\rho^2+ 6\lambda a^2\rho^2 -8a^2)}{3\rho^2 + 2M a^2}\Big]
\,.\label{Vdef}
\eea
%%%%%

   The structure of the potential $V$ can easily be studied in the
special case when $\lambda=0$, so that the background
metric is the Ricci-flat $a=b$ Myers-Perry solution \cite{myeper}.
It can be seen from (\ref{Gdef}) that in order to have an horizon, \ie
for the function $G$ to have a zero for real $\rho$, it must be that
$a^2\le \ft12 M$.  If, therefore, we write $M= 2a^2 (1+ s^2)$, where
$s$ is a real constant, then the outer horizon occurs at
%%%%%
\be
\rho_+^2 = 2a^2(1+s^2 + s\sqrt{1+s^2})\,.
\ee
%%%%%
Writing $\rho^2= \rho_+^2 + y^2$, and substituting this and $M=2a^2(1+s^2)$
into the expression (\ref{Vdef}) for the potential, one finds that
$V$ is manifestly non-negative everywhere outside the horizon, and it
tends to zero on the horizon and at infinity.

\end{document}